\documentclass[letterpaper,twocolumn,10pt]{article}
\usepackage[table,dvipsnames]{xcolor}
\usepackage{hyperref}
\usepackage{usenix2019,epsfig,endnotes}
\pdfoutput=1
\usepackage{booktabs} 
\usepackage{multirow}
\usepackage{tipa}
\usepackage{graphicx,subcaption} 
\usepackage{rotating}
\usepackage{lscape}
\usepackage{wrapfig}  
\usepackage{balance}

\newcommand{\dejavu}{{\fontfamily{cms}\selectfont dejavu}}

\begin{document}

\date{}

\title{\Large \bf Don't Pick the Cherry:\\ An Evaluation Methodology for Android Malware Detection Methods}

\author{
{\rm Aleieldin Salem}\\
Technische Universit\"{a}t M\"{u}nchen\\
salem@in.tum.de
\and
{\rm Sebastian Banescu}\\
Technische Universit\"{a}t M\"{u}nchen\\
banescu@in.tum.de
\and
{\rm Alexander Pretschner}\\
Technische Universit\"{a}t M\"{u}nchen\\
pretschn@in.tum.de
} 

\maketitle

\thispagestyle{empty}

\subsection*{Abstract}
In evaluating detection methods, the malware research community relies on scan results obtained from online platforms such as \texttt{VirusTotal}. 
Nevertheless, given the lack of standards on how to interpret the obtained data to label apps, researchers hinge on their intuitions and adopt different labeling schemes. 
The dynamicity of \texttt{VirusTotal}'s results along with adoption of different labeling schemes significantly affect the accuracies achieved by any given detection method even on the same dataset, which gives subjective views on the method's performance and hinders the comparison of different malware detection techniques.

In this paper, we demonstrate the effect of varying (1) time, (2) labeling schemes, and (3) attack scenarios on the performance of an ensemble of Android repackaged malware detection methods, called \dejavu, using over 30,000 real-world Android apps.
Our results vividly show the impact of varying the aforementioned 3 dimensions on \dejavu's performance.
With such results, we encourage the adoption of a standard methodology that takes into account those 3 dimensions in evaluating newly-devised methods to detect Android (repackaged) malware.

\maketitle

\section{Introduction}
\label{sec:introduction}
The research community has been working towards devising methods to detect Android malware (e.g., \cite{li2017automatically,pan2017dark,tian2016analysis,shahriar2015kullback}).
Despite proving to be sometimes inconsistent \cite{hurier2017euphony}, researchers rely on \texttt{VirusTotal} \cite{virustotal2019} to either download training data to evaluate their newly-devised methods \cite{wang2018beyond,suarez2017droidsieve,yang2017malware}, or to label the apps they manually gathered from the wild (e.g., app marketplaces) \cite{li2017understanding,wei2017deep,arp2014drebin}, due to the lack of better alternatives.

To be objective, \texttt{VirusTotal} does not provide users with binary labels per app, such as malicious or benign. 
It instead provides its users with the scan results of around 60 antiviral software. 
Unfortunately, such results have been found to change over time (e.g., from benign to malicious and vice versa) \cite{mohaisen2014av}. 
Furthermore, there are no standards on how to utilize the obtained scan results to label apps.
In this context, researchers use their intuition and adopt ad-hoc methods to label the apps in the datasets they train their methods with or, more importantly, release to the research community as benchmarks.
For example, based on \texttt{VirusTotal}'s scan results, Li et al.\ labeled the apps in their \emph{Piggybacking} dataset as malicious if at least one scanner deemed an app as malicious \cite{li2017understanding}, Wei et al.\ labeled apps in the \emph{AMD} dataset as malicious if 50\% or more of the total scanners labeled an app as such \cite{wei2017deep}, and the authors of the \emph{Drebin} dataset \cite{arp2014drebin} labeled an app as malicious if at least two out of ten scanners they manually selected courtesy of their reputation (e.g., AVG, BitDefender, Kaspersky, McAfee, etc.), deemed an app as malicious. 
 
Researchers have also found that training and testing a malware detection method using different datasets (i.e., constructed for different experimental attack scenarios), yields different detection results \cite{salem2018poking}. 
In fact, varying the freshness of scan results and the labeling scheme alters the distribution of malicious and benign apps in the same dataset, effectively yielding different datasets and, in turn, different detection results.
Therefore, our hypothesis is that depending on the 3-dimensions of (1) time, (2) labeling scheme, and (3) attack scenario implied by the utilized test dataset, the same malware detection method might perform differently. 
This, we argue, might hinder the comparability of different detection approaches.
Furthermore, it might incite researchers to dismiss promising detection approaches, because they underperform on a dataset with outdated labels, or because they utilize a different labeling scheme than the one adopted by the dataset's authors.

To objectively evaluate the performance of newly-devised detection methods, in this paper, we motivate the adoption of a framework that considers the dimensions of time, labeling scheme, and dataset to evaluate Android (repackaged) malware detection methods. 
To further motivate the need for such framework, we demonstrate the impact of varying the aforementioned three dimensions on the performance of the same detection method. 
In our demonstration, we focus on Android repackaged malware as a use case primarily due to the threat it poses to the Android ecosystem, namely, undermining user trust in legitimate apps, their developers, and the app distribution process. 
Using three malware datasets, viz.\ \emph{Malgenome} \cite{zhou2012dissecting}, \emph{Piggybacking} \cite{li2017understanding}, and \emph{AMD} \cite{wei2017deep} along with 1882 apps downloaded from the Google Play marketplace, we train and test a framework, \dejavu\footnote{\textbf{D\'{e}j\`{a} vu} (\textipa{/'deZA 'vy/}) is the feeling that one has experienced an event before. In essence, our framework attempts to match an app under test against a repository of previously-analyzed (i.e., seen before), apps to decide whether an app is repackaged (and potentially malicious).}, which utilizes an ensemble of detection methods that have each individually been successfully utilized by researchers to effectively detect Android repackaged\cite{zhou2012detecting, zhou2012dissecting} or piggybacked \cite{li2017understanding, li2017automatically} malware. 
Such methods include compiler fingerprinting, probabilistic classification using a naive Bayes classifier, and matching apps according to their metadata, components, classes, and methods. 


Our results show that the scan results obtained from \texttt{VirusTotal} are continuously changing with time and, hence, should not be taken for granted unless they are up-to-date.
Secondly, we found that such \texttt{VirusTotal} results--regardless of their freshness--can significantly alter the composition of a dataset (i.e., which apps are malicious, and which are benign), depending on the scheme adopted to label apps in a dataset. 
This phenomenon affected the detection accuracy of our ensemble method, \dejavu, by 37.5\% (i.e., 0.72 versus 0.99), which gives different views of the method's effectiveness. 

\textbf{The contributions} of this paper are:
\begin{itemize}
    \setlength\itemsep{0em}
    \item A malware detection evaluation methodology, which mandates varying the freshness of scan results (e.g., from \texttt{VirusTotal}), the scheme adopted to label apps, and the attack scenario (section \ref{sec:evaluation}), and a demonstration of the effect of varying the aforementioned dimensions on the detection accuracies of an ensemble of malware detection methods called \dejavu\ (described in section \ref{sec:dejavu}). 
    \item An analysis of the samples that are labeled differently when using different labeling schemes, which concludes that the majority of such apps are \texttt{Adware} and the majority of the URL's they contact are benign (section \ref{sec:discussion}). 
    \item We make the results of our experiments and \dejavu's source code available online \footnote{https://goo.gl/p7hkH9}. 
\end{itemize}

\noindent
In section \ref{sec:background}, we motivate the need for our research using examples that we encountered during the implementation 
\dejavu, and briefly discuss the attack scenarios usually encountered by Android repackaged malware detection methods.
Section \ref{sec:threats} discusses threats to validity.
Related work is presented in section \ref{sec:related} and the paper is concluded in section \ref{sec:conclusion}.

\section{Background}
\label{sec:background}
In this section, we motivate the need for our paper and its line of research with examples. 
The first example in section \ref{sec:motivation} demonstrates the importance of considering the dimensions of time and labeling scheme.
In section \ref{subsec:attack_scenarios}, we give examples of attack scenarios under which a detection method can encounter Android repackaged malware. 
\subsection{Motivating Example}
\label{sec:motivation}
During our evaluation of \dejavu's components, particularly the \emph{quick matching} module, against the \emph{Piggybacking} dataset we came across multiple dubious scenarios.
In summary, the \emph{quick matching} module utilizes compiler fingerprinting to classify apps as malicious or benign if their codebases do not match. 
One of the scenarios we encountered was during testing an app called \texttt{TP.LoanCalculator}. 
Despite being labeled by the authors of the \emph{Piggybacking} dataset as malicious, the \emph{quick matching} module deemed the test app\footnote{2b44135f245a2bd104c4b50dc9df889dbd8bc79b} as benign. 
After inspection, we found that the module classified the aforementioned test app as benign because it matched one\footnote{d8472cf8dcc98bc124bd5144bb2689785e245d83} of the apps \dejavu\ keeps as a reference dataset (section \ref{sec:dejavu}) in terms of metadata, used compiler, and even codebase. 

Given that \emph{Piggybacking} comprises benign apps and their repackaged malicious versions, it is expected to find apps that match in terms of metadata, graphical user interface, and file names. 
However, two apps possessing the exact same codebase implies that they are basically the same app, perhaps with slight modifications to their resource files (e.g., strings and colors). 
So, how can two apps in the same dataset and with identical codebases be simultaneously deemed as malicious and benign?
One possible answer to this question is that the malicious app altered the resource files of the original one (e.g., to change a URL), which we found not to be the case.
The more plausible answer is that either, or both, apps were given incorrect labels. 

In \cite{li2017understanding}, the authors of the \emph{Piggybacking} dataset labeled the apps they gathered with the aid of \texttt{VirusTotal} scan results. 
So, we retrieved the \texttt{VirusTotal} results of both apps by querying the platform's web interface.
The test app was labeled malicious by 14 out of 60 antiviral software scanners. 
We noticed, however, that the results acquired from \texttt{VirusTotal} indicated that the app was last analyzed in 2013. 
So, we uploaded the app's APK archive for re-analysis in late 2018 to see whether the number of scanners would differ. 
The 2018 version of the scan results indicated that three more scanners deemed the app malicious. 

The scan results obtained for the reference app were more interesting. 
Similar to its malicious version, this presumably benign app was last analyzed in 2013. 
In this case, the scan results from \texttt{VirusTotal} indicated that all scanners deemed the app as benign, displaying green "\textbf{Clean}" labels next to the names of all scanners.
Nevertheless, after clicking on the \textbf{Reanalyze} button, 17 of the green labels turned red displaying different malware family names that indicate the malignancy of the app.
That is to say, the reference app originally labeled and released as part of the \emph{Piggybacking} dataset as a benign app is, in fact, another version of a malicious app of the type \texttt{Adware}. 

Needless to say, the authors of \emph{Piggybacking} did not intentionally mislabel apps. 
The most likely scenario is that, at the time of releasing the dataset, the reference app was still deemed as benign by the \texttt{VirusTotal} scanners.
Our example shows the evolution of scan results returned by \texttt{VirusTotal} and the impact of time on them. 
This phenomenon implies that prior to using any datasets--including manually-labeled ones, all apps need to be re-analyzed and re-labeled. 

Having mentioned labeling our example shows that 17 out of 60 scanners (i.e., 28.33\% of scanners), deemed both apps as malicious. 
Interestingly, some renowned scanners including AVG, McAfee, Kaspersky, Microsoft, and TrendMicro continue to deem both apps as benign. 
However, according to the authors' criterion to label an app as malicious if at least \textbf{one} scanner deems it so \cite{li2017understanding}, both apps would be labeled as malicious.
The same would not hold for the authors of the \emph{AMD} dataset who consider an app as malicious if at least 50\% of the \texttt{VirusTotal} scanners deem it malicious \cite{wei2017deep}. 
Those two dimensions (i.e., time and labeling scheme/criteria), are expected to affect the performance of any detection framework using the same dataset, especially if the malicious apps are repackaged or belong to ambiguous malware types such as \texttt{Adware} or \texttt{Riskware}.
This paper attempts to demonstrate the effect of varying those two dimensions on the performance of an ensemble method, \dejavu, built to detect Android repackaged malware. 
\subsection{Attack Scenarios}
\label{subsec:attack_scenarios}
Consider a malware author (hereafter attacker) who wishes to write a malicious app that (1) resembles a renowned, benign app to trick users into downloading what they believe is a new version of a popular app, and (2) manages to evade detection by any app vetting mechanisms employed by the marketplace(s) the attacker targets. 
To achieve the first objective, the attacker can repackage a benign app with malicious content which is, in fact, a straightforward task \cite{salem2018repackman, li2017understanding}. 

As for the second objective, researchers have identified two different scenarios that can be adopted by attackers to evade detection by vetting mechanisms employed by app marketplaces \cite{salem2018poking,wang2018beyond}. 
In both scenarios, we assume that the attacker does not have access to information about (a) the methods (or lack thereof) used by the target marketplace to vet apps, and (b) the dataset of apps used to train the marketplace's vetting mechanism. 
By training, we do not assert the use of machine learning classifiers; instead, we refer to apps used by a vetting mechanism as references of what is malicious and what is benign. 
The only information we assume the attacker has access to is the list of apps hosted on a marketplace, which can be easily obtained (e.g., by crawling a marketplace) \cite{wang2018beyond}. 
Given such limited information, the attacker might assume that marketplaces use the apps they host as references of benign apps, especially since such apps have undergone a vetting process upon being uploaded. 
Based on this assumption, we identify two scenarios: \emph{conventional} and \emph{confusion}, which we describe in the following paragraphs.

\paragraph{\textbf{Conventional Scenario}}
On the one hand, the attacker may opt to upload the repackaged app ($\alpha^{*}$) to a marketplace ($\mu$) that does \textbf{not} host its original, benign version ($\alpha$). 
This means that the vetting mechanism employed by ($\mu$) will be faced with an out-of-sample app (i.e., one it has never seen before), and hence cannot easily match to either benign or malicious apps. 
Since it has been found that the majority of repackaged apps dwell on marketplaces on which their original versions are not hosted \cite{wang2018beyond}, we refer to this common scenario as \emph{conventional}.

\paragraph{\textbf{Confusion Scenario}}
On the other hand, the attacker may adopt the opposite approach by opting to upload their repackaged app ($\alpha^{*}$) to a marketplace ($\mu$) on which the original app ($\alpha$) is hosted. 
The rationale behind this choice is that vetting mechanisms can be perplexed by ($\alpha^{*}$) which resembles a benign app it has seen before. 
This scenario is indeed the less common \cite{wang2018beyond}, yet has been found to be more effective against some techniques, such as machine learning classifiers \cite{salem2018poking}. 

For example, an attacker targeting Google Play might decide upon downloading a popular gaming app (e.g., \texttt{com.rovio.angrybirds}), grafting it with a payload that displays advertisements to the users, slightly altering its appearance (e.g., via modifying colors or strings), and re-uploading it to Google Play under a slightly different name (e.g., \texttt{com.rovio.crazybirds}).
If Google Play indeed uses \texttt{com.rovio.angrybirds} in training its app vetting mechanism, it is likely that such mechanism deems \texttt{com.rovio.crazybirds} as benign, especially since the repackaged app's components, structure, and even behavior are expected to match those of its original version.

\section{dejavu}
\label{sec:dejavu}
\begin{figure}[!t]
\centering
\includegraphics{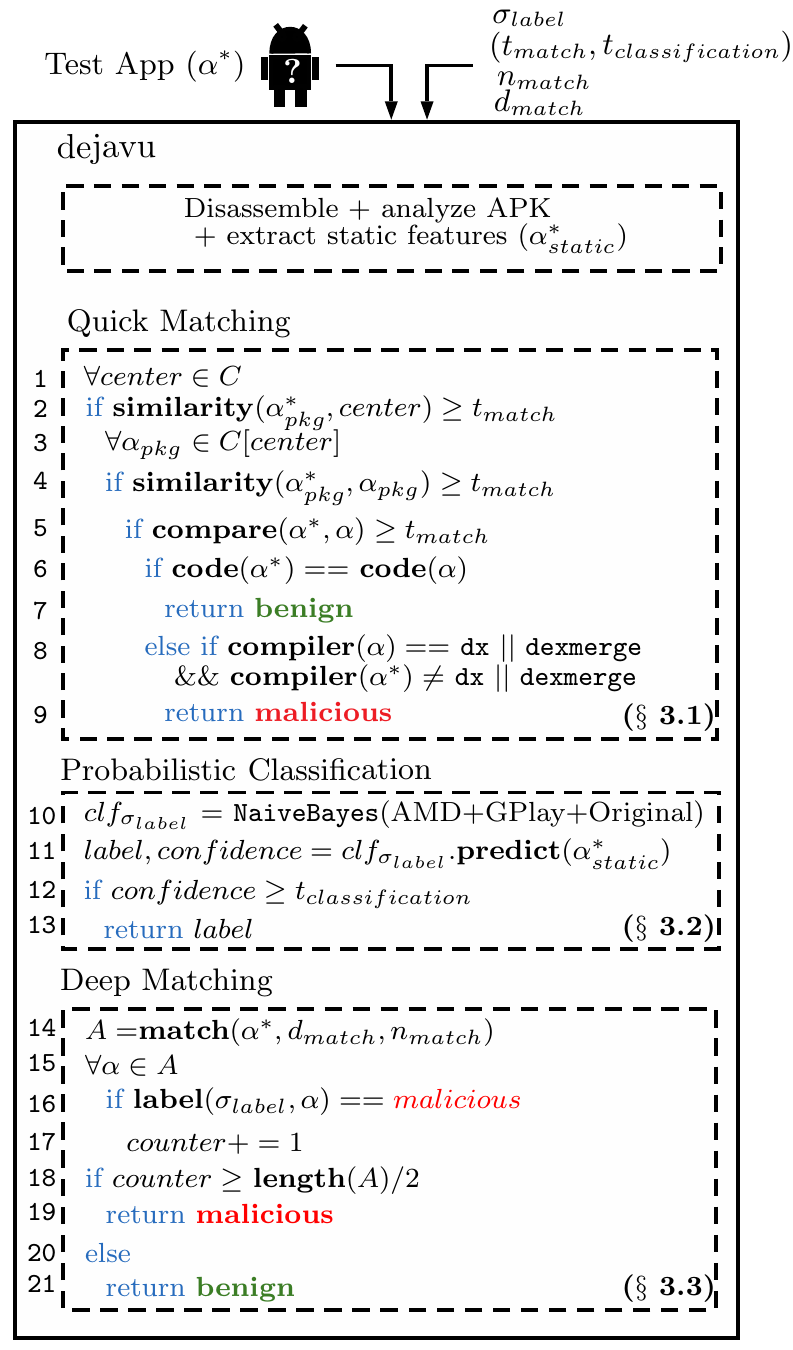}
\caption{An overview of the process adopted by \dejavu\ to deem Android apps as malicious or benign. The framework implements three types of detection methods (dashed boxes) that are executed in sequence (i.e., top-down). The numbers on the bottom right of every box depict the section number in which the corresponding detection method is discussed.}
\label{fig:overview}
\end{figure}

We implemented \dejavu\ to utilize three different methods that have been successfully used to detect Android repackaged malware. 
Each detection method is implemented as a separate module, viz. \emph{quick matching}, \emph{probabilistic classification}, and \emph{deep matching}.
As seen in figure \ref{fig:overview}, the primary input to \dejavu\ is an Android test app ($\alpha^{*}$) which is sequentially tested by \dejavu's methods to decide upon its class (i.e., malicious or benign).

To emulate the attack scenarios we discussed in section \ref{subsec:attack_scenarios}, we implemented \dejavu\ to resemble a marketplace's vetting mechanism making use of a repository of Android apps that depict a reference dataset against which test apps are compared. 
To hasten \dejavu's decision about an app's class, apps in the reference dataset have been previously decompiled and analyzed. 
The data extracted from each app is saved to a directory which contains a dictionary of information about the app's components (e.g., lists of activities, files, libraries, services, etc.), the app's icon, and a vector of static features extracted from its APK archive (listed in appendix \ref{appendix:static_features}). 

The labeling scheme to be adopted during classification is dictated by the parameter ($\sigma_{label}$) and can have the following values:
\begin{itemize}
 \setlength\itemsep{0em}
 \item \emph{vt1-vt1} labels an app malicious if one or more \texttt{VirusTotal} scanners deem it as malicious; otherwise, it is labeled as benign. 
 \item \emph{vt50p-vt50p} labels an app malicious if 50\% or more of the \texttt{VirusTotal} scanners report it as malicious; otherwise, the app is labeled as benign. 
 \item \emph{vt50p-vt1} labels an app malicious if 50\% or more of the scanners deem it malicious and labels an app as benign if none of such scanners deem it malicious. 
\end{itemize}

Not only does the value of ($\sigma_{label}$) affect the ground truth label of ($\alpha^{*}$), it instructs \dejavu\ to use different sets of apps in the reference dataset, viz.\ the ones that pass the criteria of the labeling scheme, to match apps and train classifiers. 
In other words, the value of ($\sigma_{label}$) can alter the composition of, both, the reference dataset and test apps. 

\subsection{Quick Matching}
\label{subsec:quick_matching}
The \emph{quick matching} detection method uses three techniques to classify the test app ($\alpha^{*}$) as quickly and as reliably as possible, prior to delving into more time-consuming detection methods. 
Firstly, it checks whether the test app ($\alpha^{*}$) matches any benign app(s) in \dejavu's reference dataset. 
If a match is found, the method uses the techniques of codebase comparison and compiler fingerprinting to decide upon ($\alpha^{*}$)'s class. 
%
\paragraph{\textbf{Repackaging Detection}}
To trick users into voluntarily downloading their repackaged malicious apps, malware authors might opt to retain some basic information (or metadata) of the original benign app, such as app package name (e.g., \texttt{com.my.app}), app description, icon, and so forth.
This behavior can be used to match test apps to benign ones in our reference dataset, which facilitates making decisions about their nature. 

To rapidly figure out whether ($\alpha^{*}$) can be matched using \emph{quick matching}, we attempt to match its package name to those of benign apps (i.e., according to $\sigma_{label}$), in \dejavu's reference dataset.
The package names of such benign apps are clustered--using the Levenshtein distance as a distance metric \cite{shrivastava2013text}--and saved as a dictionary ($C$); the values in the dictionary are lists of app package names, whereas the keys are the centers of each cluster.
For example, the cluster of apps $\lbrace$\texttt{jp.colopl.allenCN}, \texttt{jp.colopl.enkare}, \texttt{jp.colopl.krarmy}, $\dots\rbrace$ has the center \texttt{jp.colopl.entrain}.

Firstly, the package name of ($\alpha^*$), denoted ($\alpha^{*}_\mathit{pkg}$), is compared against the centers ($center$) of different clusters ($C$) (lines \texttt{1} and \texttt{2} in figure \ref{fig:overview}).
The comparison is based on the Jaccard index \cite{cohen2003comparison}, which returns a number between $0.0$ and $1.0$. 
If the similarity between ($\alpha^{*}_\mathit{pkg}$) and ($center$) is greater than or equal to a matching threshold ($t_\mathit{match}$) (i.e., line \texttt{2}), the cluster is investigated further.
That is to say, every package name in the cluster ($\forall \alpha_\mathit{pkg}\in center$) is compared against the test app's ($\alpha^{*}_{pkg}$) (line \texttt{3}). 
In line \texttt{4}, if the package name of ($\alpha^{*}$) matches that of an app ($\alpha$) in the reference dataset with similarity greater than or equal to ($t_{match}$), the metadata (i.e., package names, descriptions, and icons), of both apps are compared (line \texttt{5}).
If such metadata is similar with threshold greater than the ($t_\mathit{match}$) threshold, we assume that ($\alpha^{*}$) is some version of ($\alpha$). 
To decide whether ($\alpha^{*}$) is malicious or benign, \emph{quick matching} uses the techniques of codebase comparison and compiler fingerprinting. 
\paragraph{\textbf{Codebase Comparison}}
Regardless of being malicious or benign, if ($\alpha^{*}$) shares the exact codebase with ($\alpha$), then we assume that both apps are effectively the same in terms of functionality.     
It follows that, if ($\alpha$) is labeled as benign according to ($\sigma_{label}$), then ($\alpha^{*}$) should be deemed as benign as well. 
The question is: what if ($\alpha^{*}$) is, in fact, a repackaged version of ($\alpha$) grafted with a malicious payload. 
If that is true, then the attacker needs to alter ($\alpha$)'s codebase to add their malicious code or the code that dynamically loads it, which contradicts the assumption above.
In this context, we implemented \emph{quick matching} to compare the codebases of ($\alpha^{*}$) and ($\alpha$) (i.e., line \texttt{6}), and deem the former app as benign if the codebases are identical.

In checking the codebase, \dejavu's \emph{quick matching} detection method compares the \texttt{classes.dex} files of both apps, and ignores the resource files. 
However, we believe that this does not affect the check's effectiveness. 
The only scenario we could think of that could render the codebase comparison technique ineffective is if the original app ($\alpha$) stores \texttt{Java} or \texttt{Dex} bytecode as a string that is stored in the resource files (e.g., \texttt{strings.xml}), and dynamically loads it during runtime. 
In this case, the attacker can modify this code to include their malicious code and repackage the app.
Despite its technical possibility, we argue that such scenario is highly unlikely to manifest, especially since Android developers have more advanced, reliable, and secure methods to dynamically alter the functionalities and appearances of their apps (e.g., via App Bundles \cite{android2019appbundle}).
\paragraph{\textbf{Compiler Fingerprinting}}
There are some scenarios in which the codebases of ($\alpha^{*}$) and ($\alpha$) could differ, yet both apps could be benign. 
For example, ($\alpha^{*}$) may be an updated version of ($\alpha$), which indeed should have a different codebase. 
However, the test app ($\alpha^{*}$) may be a repackaged version of ($\alpha$) withholding malicious code that uses the original codebase as a facade. 
To differentiate between such scenarios, \emph{quick matching} uses the technique of compiler fingerprinting proposed by Stazzere\cite{stazzere2016detecting}, which detects Android repackaged malware based on the following assumption: legitimate developers usually have access to their apps' source code and, hence, their apps should be compiled using the \texttt{dx} or \texttt{dexmerge} compilers that ship with the Android SDK. 
Consequently, apps that are compiled using third-party compilers used by reverse engineering tool (e.g., \texttt{dexlib}), should raise major suspicions.

Based on this idea, \emph{quick matching} fingerprints ($\alpha^{*}$)'s and ($\alpha$)'s compilers using {\fontfamily{lms}\selectfont APKiD} \cite{apkid2018} and compares them (i.e., line \texttt{8}).  
Given that it has been already established that the codebases of ($\alpha^{*}$) and ($\alpha$) are different, the scenario in which ($\alpha$)'s compiler is \texttt{dx} or \texttt{dexmerge}, whereas ($\alpha^{*}$)'s compiler is a third-party compiler (e.g., \texttt{dexlib}), implies that the developer of ($\alpha^{*}$) most likely did not have access to the original app's source code and had to repackage it. 
If this condition holds, \emph{quick matching} deems ($\alpha^{*}$) malicious, as seen in line \texttt{9}.

One might argue that some developers may elect to decompile and repackage their apps to include slight modifications (e.g., fix a typographical error), without having to recompile the source code, which should undermine the correctness of the check in line \texttt{8}. 
We argue that such slight modifications should not affect the app codebase.
So, even if ($\alpha^{*}$) embodies a repackaged version of ($\alpha$), having the same codebase should help deem it benign according to the check in line \texttt{6}.

One exception to this general case is if legitimate developers elect to alter their codebases (e.g., to change a constant \texttt{String}), on the \texttt{smali} level and repackage their apps. 
However, legitimate developers that possess their apps' code have no good reasons to work on the \texttt{smali} level to alter their codebases. 
Given the rarity of this scenario and the severity of false negatives, in this case, we opt to deem test apps that pass the check in line \texttt{8} as malicious.
\paragraph{\textbf{Other Scenarios}}
The checks used between lines \texttt{1} and \texttt{9} do not cover all possible scenarios. 
Firstly, there are the scenarios in which \emph{quick matching} was not able to match ($\alpha^{*}$) to any apps in \dejavu's reference dataset. 
Secondly, there are scenarios in which the codebases of ($\alpha^{*}$) and ($\alpha$) are not the same, yet they share the same compilers. 
For example, if both ($\alpha^{*}$) and ($\alpha$) are compiled using the \texttt{dx} compiler, that could either mean that (a) ($\alpha^{*}$) is a benign update to the benign app ($\alpha$), or (b) ($\alpha^{*}$) is a malicious update to a benign app ($\alpha$) whose source code is available online (e.g., K-9 Mail \cite{k9mail2019}). 
In those cases, \emph{quick matching} cannot make confident decisions about ($\alpha^{*}$)'s class and defers the making of this decision to \emph{probabilistic classification}. 
\subsection{Probabilistic Classification}
\label{subsec:probabilistic_classification}
The inability of \emph{quick matching} to classify ($\alpha^{*}$) with absolute confidence according to domain-knowledge-based assumptions implies uncertainty.
To deal with such uncertainty, we widen the scope within which we attempt to match the test app ($\alpha^{*}$).
So, instead of matching the test app ($\alpha^{*}$) to a specific app in terms of metadata, we attempt to match it to a class of apps (i.e., malicious or benign), using information about the files in its APK archive, its components, the permissions it requests, and the API calls it utilizes. 
Furthermore, instead of yielding absolute confidence about the test app ($\alpha^{*}$)'s class, we relax this requirement and return a probabilistic measure. 
Such probabilistic measure, we argue, suits types and families of malware that are ambiguous (e.g., \texttt{Adware}), which are considered to be potentially unwanted rather than purely malicious. 
Lastly, we still aspire to maintain the efficiency of the classification process by classifying a test app as quickly as possible.

To achieve such objectives, the \emph{probabilistic classification} detection method relies on a probabilistic machine learning classifier to infer the class of ($\alpha^{*}$) for the following reasons. 
In general, machine learning proved to be a reliable method in the field of (Android) malware detection and has, hence, been utilized by various researchers within this domain \cite{mariconti2017mamadroid,suarez2017droidsieve,arp2014drebin}. 
Furthermore, once features are extracted from Android apps, machine learning classifiers are usually faster to train and validate than more complex methods (e.g., dependency graph isomorphism \cite{crussell2015andarwin}). 
However, the abundance of research in this field and the utilization of different datasets and features make it difficult to compare their performances to decide upon a few approaches to consider \cite{pendlebury2018enabling}. 
Consequently, we chose a classifier that yields classification confidence in the form of a probability along with the class label. 
 
The classifier we use is a multinomial Naive Bayes classifier trained using static features extracted from the apps in \dejavu's reference dataset. 
Those static features, enumerated in appendix \ref{appendix:static_features}, depict information about the apps' components \cite{sato2013detecting,sanz2013puma}, the permissions they request \cite{arp2014drebin}\cite{wu2012droidmat}, the API calls found in their codebases \cite{zhou2013fast}, and the compilers used to compile them.

According to the value of ($\sigma_{label}$), the classifier will be trained using the apps in the reference dataset which satisfy the labeling scheme's criteria (i.e., line \texttt{10}); hence, the name ($clf_{\sigma_{label}}$).  
For example, if the labeling scheme is \emph{vt50p-vt1}, the classifier will be trained using malicious apps deemed by more than 50\% of \texttt{VirusTotal} scanners and using benign apps deemed malicious by no \texttt{VirusTotal} scanners. 

To classify ($\alpha^{*}$), the aforementioned static features are extracted from the app to yield a vector of numerical features ($\alpha^{*}_{static}$). 
In line \texttt{11}, the classifier ($clf_{\sigma_{label}}$) uses ($\alpha^{*}_{static}$) to classify the test app and returns a ($label$) corresponding to the class (i.e., malicious or benign), it is more confident ($\alpha^{*}$) belongs to along with a ($confidence$) in such decision in the format of a probability. 
If the returned ($confidence$) is greater than or equal to a threshold passed to \dejavu, called ($t_{classification}$), the \emph{probabilistic classification} method returns the ($label$) (lines \texttt{12-13}). 
Otherwise, the task of classifying ($\alpha^{*}$) is further deferred to the \emph{deep matching} detection method. 
\subsection{Deep Matching}
\label{subsec:deep_matching}
The last detection method used by \dejavu\ is \emph{deep matching}. 
Unlike \emph{quick matching}, the \emph{deep matching} method attempts to match the test app ($\alpha^{*}$) to a number of similar malicious or benign apps in the reference dataset. 
The method defines the similarity between apps in terms of the overlap between different app information. 
For example, given two lists of strings, say ($L_1$) and ($L_2$), depicting the files included in the APK archives of two apps, \emph{deep matching} defines similarity in terms of the Jaccard index between the lists (i.e., $\frac{|L_1\cap L_2|}{|L_1\cup L2|}$). 

In line \texttt{14}, \emph{deep matching} takes the test app ($\alpha^{*}$) as an input along with an analysis depth ($d_{match}$) and an upper bound on the apps to match to ($\alpha^{*}$). 
The upper bound parameter, ($n_{match}$), limits the number of apps against which ($\alpha^{*}$) is allowed to be matched. 
This parameter is meant to speed up the matching process by instructing the \textbf{match} method to conclude its matching process once it finds ($n_{match}$) apps that are similar to ($\alpha^{*}$) with a threshold of at least $0.67$ (i.e., two-thirds), and returns the matched apps as a set ($A$).

The exact information used by the \emph{deep matching}'s \textbf{match} method (to compare and match apps), depends on the value of the ($d_{match}$) parameter, which depicts the \emph{depth} of the extracted information.
If $d_{match}==1$, \emph{deep matching} compares the apps' metadata in a manner similar to how \emph{quick matching} compares ($\alpha^{*}$) and ($\alpha$) in line \texttt{5}, viz.\ the apps' package names, descriptions, and the structural similarity of their icons according to the SSIM measure \cite{wang2004image}.
If $d_{match}==2$, the method retrieves lists of all components declared by the two apps being compared and calculates the Jaccard index between them. 
To avoid prolonging the matching process, \emph{deep matching} does not compare the content of components; instead their names.
Lastly, if the value of $d_{match}==3$, the list of resource files, libraries, classes, and methods are compared using the Jaccard distance as well. 

The \textbf{match} method is implemented in a cumulative manner that includes similarity scores from depth levels lower than the level passed to the method in the final similarity score. 
For example, if ($d_{match}==3$), the method would include the similarity score achieved at depths one, two, and three in the final similarity score it returns. 
The overall similarity score between two apps is an unweighted average of all the scores returned by each depth level. 
For instance, if matching depths one, two, and three returned scores of $0.8$, $0.65$, and $0.79$ the final score returned will be $0.74$. 
So, the higher the value of ($d_{match}$), the more the information retrieved and compared.

For each app ($\alpha$) matched by \textbf{match}, the label of this app is retrieved in accordance with the labeling scheme ($\sigma_{label}$) passed to \dejavu. 
The method keeps track of the number of apps labeled as malicious (lines \texttt{16-17}). 
The final verdict returned by \emph{deep matching} vis-\`{a}-vis ($\alpha^{*}$)'s class is determined as a simple majority vote of the labels/classes of the matched apps (i.e., $50\%+1$).
For instance, if ($\alpha^{*}$) is matched to a set of ten apps ($A$) six of which are malicious, the app ($\alpha^{*}$) will be labeled as malicious (line \texttt{19}).
The test app ($\alpha^{*}$) will be labeled as benign (line \texttt{21}) if 50\% or less of the apps in ($A$) are malicious, according to ($\sigma_{label}$). 

\section{Evaluation}
\label{sec:evaluation}
In this section, we use the detection methods of \dejavu\ to investigate the effect of varying the dimensions of time, labeling schemes adopted by researchers to label apps in their datasets, and attack scenarios on the detection accuracies of such methods. 
To tackle such concerns, we postulate the following research questions and devise experiments that address them:
\begin{itemize}
    \setlength\itemsep{0.1em}
    \item[\textbf{RQ1}] How do labeling schemes affect the choice of \dejavu's input parameter values?
    \item[\textbf{RQ2}] How does the variation of label values in \textbf{time}, affect the accuracies of detection methods?
    \item[\textbf{RQ3}] How does varying the \textbf{labeling scheme} affects the accuracies of detection methods?
    \item[\textbf{RQ4}] How do detection methods perform across different \textbf{scenarios} (i.e., \emph{conventional} versus \emph{confusion})?
    \item[\textbf{RQ5}] What effect does combining detection methods have on detection performance?
    \item[\textbf{RQ6}] What do the apps that do not fit either criteria in the \emph{vt50p-vt1} labeling scheme comprise?
\end{itemize}
\subsection{Datasets}
\label{subsec:datasets}
\bgroup
\def\arraystretch{1.0}
\begin{table*}[]
\caption{The impact of time and labeling schemes on the composition of Android datasets.}
\label{tab:reference}
\footnotesize
\resizebox{\textwidth}{!}{
\begin{tabular}{|c|c|c|c|c|c|c|}
\hline
\cellcolor[HTML]{C0C0C0} & \multicolumn{4}{c|}{\cellcolor[HTML]{C0C0C0}\textbf{Labeling Scheme} ($\sigma_{label}$)} & \cellcolor[HTML]{C0C0C0} & \cellcolor[HTML]{C0C0C0} \\ \cline{2-5}
\multirow{-2}{*}{\cellcolor[HTML]{C0C0C0}\textbf{Dataset}} & \cellcolor[HTML]{EFEFEF}\emph{original} & \cellcolor[HTML]{EFEFEF}\emph{vt1-vt1} & \cellcolor[HTML]{EFEFEF}\emph{vt50p-vt50p} & \cellcolor[HTML]{EFEFEF}\emph{vt50p-vt1} & \multirow{-2}{*}{\cellcolor[HTML]{C0C0C0}\begin{tabular}[c]{@{}c@{}}\textbf{More Malicious}\\ \textbf{Scan Results}\end{tabular}} & \multirow{-2}{*}{\cellcolor[HTML]{C0C0C0}\begin{tabular}[c]{@{}c@{}}\textbf{Less Malicious}\\ \textbf{Scan Results}\end{tabular}} \\ \hline 
\emph{Piggybacked} (2014-2017) & 1399 (malicious) & \textbf{1263 (90\%)} & 159 (11.36\%) & 159 (11.36\%) & 845 (60.4\%) & 234 (16.72\%) \\ \hline
\emph{Original} (2014-2017) & 1355 (benign) & \textbf{852 (63\%)} & 1348 (99.5\%) & 852 (63\%) & 219 (16.16\%) & 212 (15.64\%) \\ \hline
\emph{Malgenome} (2010-2012) & 1234 (malicious) & \textbf{1234 (100\%)} & 1234 (100\%) & 1234 (100\%) & 366 (29.65\%) & 609 (49.35\%) \\ \hline
\emph{Gplay} (2017) & 1882 (benign) & \textbf{1572 (83\%)} & 1837 (99.46\%) & 1572 (83\%) & 107 (5.7\%) & 79 (4.2\%) \\ \hline
\emph{AMD} (2010-2016) & 24553 (malicious) & 24552 (99.9\%) & \textbf{12765 (52\%)} & 12765 (52\%) & 12481 (50.83\%) & 6454 (26.3\%) \\ \hline
\end{tabular}
}
\end{table*}
\egroup

In this section, we briefly discuss the composition of the datasets we used to train and test \dejavu.
The largest, and most recent, dataset we use is \emph{AMD} \cite{wei2017deep}, which comprises 24,553 malicious apps of different malware families and types. 
The dataset provides us with a comprehensive view of malicious behaviors that can be found in Android malware.
Consequently, \dejavu\ uses the apps from the \emph{AMD} dataset as references to malicious behaviors and includes them in its reference dataset.

To complement the malicious apps in \dejavu's reference dataset with the benign ones, we downloaded around 1900 apps from the Google Play store and added them to the reference dataset. 
We do not assert that such apps are benign, given the unfortunate fact that some malicious apps make it past Google Play's app vetting mechanisms and dwell on the marketplace \cite{wang2018beyond}. 
Instead, all apps in the reference dataset are dynamically labeled according to the specified labeling scheme ($\sigma_{label}$), and based on recently downloaded \texttt{VirusTotal} scan reports, prior to being used to train any detection methods. 

We use two datasets to test \dejavu's detection methods emulating the \emph{conventional} and \emph{confusion} scenarios discussed in section \ref{subsec:attack_scenarios}. 
The first test dataset we utilize is \emph{Malgenome} \cite{zhou2012dissecting}.
It originally comprised 1260 malicious apps, almost 86\% of which were found to be repackaged malware instances. 
Despite being released in 2012, malicious apps that were originally released as part of \emph{Malgenome} continue to exist in Android app marketplaces and, consequently, in more recent Android malware datasets \cite{salem2018poking, wei2017deep}.
Using this dataset, we wish to assess \dejavu's ability to successfully recognize malicious apps that do not match to any benign ones used by the framework (i.e., \emph{conventional} scenario).

The second test dataset we use is \emph{Piggybacking} \cite{li2017understanding}. 
The dataset comprises 1400 pairs of original apps along with their repackaged versions gathered between 2014 and 2017.
Given the hashes of the apps belonging to this dataset, we downloaded as many apps as possible from the \emph{Androzoo} repository \cite{allix2016androzoo}. 
We managed to acquire 1355 original, benign apps and 1399 of their repackaged, malicious versions.
The reason behind such an imbalance is that some original apps have more than one repackaged version.
In our experiments, we refer to the benign segment of the dataset as \emph{Original}, and the dataset's malicious segment as \emph{Piggyback\textbf{\underline{ed}}} (not to be confused with \emph{Piggybacking}, which also includes the \emph{Original} segment).
We use this dataset to simulate the \emph{confusion} scenario in which an attacker targets a marketplace with a repackaged malware whose benign version is hosted by the marketplace.
So, we include the \emph{Original} segment in \dejavu's reference dataset and test the framework's detection methods using both, the \emph{Original} and \emph{Piggybacked} segments.
The former segment of the dataset is used to ensure that the techniques adopted by \dejavu's detection methods do not prevent it from correctly classifying benign apps it has seen before.
The latter segment depicts the manifestation of the \emph{confusion} scenario.

\subsection{Dataset Composition Analysis}
As discussed in section \ref{sec:introduction}, adopting different labeling schemes alters the composition of the same dataset. 
In table \ref{tab:reference}, we show the impact of, both, time and labeling schemes on each of the datasets enumerated in section \ref{subsec:datasets}. 
For each \textbf{Dataset}, the table shows the number of malicious/benign apps the dataset contained at the time of release (i.e., using the \emph{original} labeling scheme). 
The emboldened values in the \textbf{Labeling Scheme ($\sigma_{label}$)} column, depicts the same labeling scheme adopted by the dataset authors to label apps. 
For example, malicious apps in the \emph{Piggybacked} dataset were labeled as such if at least one \texttt{VirusTotal} scanners deemed them malicious. 
However, the numbers under \emph{vt1-vt1}, \emph{vt50p-vt50p}, and \emph{vt50p-vt1} are according to the latest \texttt{VirusTotal} scan results that we acquired between December 2018 and January 2019 after re-analyzing all apps.

Comparing the numbers in the \emph{original} column and the emboldened ones gives us an indication of how time affects the composition of each dataset. 
As seen in the table, apart from the \emph{Malgenome} dataset, time has a substantial effect on the number of apps that continue to be labeled as malicious or benign according to the labeling scheme adopted by the dataset's authors at the time of release. 
For example, in \cite{li2017understanding}, the \emph{Piggybacked} segment of the \emph{Piggybacking} dataset comprised 1399 apps that had \textbf{at least one} \texttt{VirusTotal} scanner deeming them as malicious. 
Using the same criterion, as of early 2019, about 10\% of such apps seize to be labeled as malicious and switch to becoming benign.  
The effect of time on the \emph{Original} segment of the aforementioned dataset is more significant; about 38\% of the apps originally labeled as benign (i.e., with no \texttt{VirusTotal} scanners deeming them malicious), were updated to be malicious. 
Similarly, 17\% of the apps we downloaded from the Google Play marketplace, which are presumed to have undergone rigorous vetting prior to being available for download, were deemed malicious by at least on \texttt{VirusTotal} scanner. 
The same effect of time can be noticed on the \emph{AMD} dataset where almost 50\% of its apps failed to pass the labeling criterion its authors used to deem an app malicious (i.e., at least 50\% of \texttt{VirusTotal} scanners deem an app as malicious). 

As part of studying the effect of time on the labels of apps in Android (malware) datasets, we studied the difference in the number of \texttt{VirusTotal} scanners that deemed apps as malicious/benign between two scans, where we triggered the latest scan of apps in the aforementioned datasets between November 2018 and January 2019. 
Each scan report downloaded from \texttt{VirusTotal} includes a field, \emph{positives\_delta}, that depicts that difference as a signed integer. 
A positive difference indicates an increased number of scanners deeming an app malicious, whereas a negative difference indicates a decreased number of such scanners. 
In table \ref{tab:reference}, we summarize those differences for all datasets.
The fields (\textbf{More Malicious Scan Results}) and (\textbf{Less Malicious Scan Results)} show the number of apps in each dataset that had the number of \texttt{VirusTotal} scanners deeming them as malicious, respectively, increase and decrease. 
The time interval between two scans might differ from one app to another. 
So, the numbers under both columns do not depict the change in \texttt{VirusTotal} scan results over a fixed period of time.
However, such results show that the scan results of \texttt{VirusTotal} are dynamic and continuously changing, which directly affects the composition of each dataset. 
That is to say, each time an app is (re)analyzed, depending on the labeling scheme, it might have a different number of \texttt{VirusTotal} scanners deeming it as malicious and, thus, may switch class from being malicious to benign or vice versa. 
Such volatility can render decent results achieved by a given detection method less indicative of the method's quality, especially since it has been achieved on an obsolete, perhaps incorrect, labeling.

\bgroup
\def\arraystretch{1.0}
\begin{table*}[]
\caption{A summary of the detection accuracies achieved by \dejavu's detection methods individually and as an ensemble under different labeling schemes.
The labeling schemes \emph{original}, \emph{vt1-vt1}, \emph{vt50p-vt50p}, and \emph{vt50p-vt1} are shortened to \textbf{org}, \textbf{vt11}, \textbf{vt50}, and \textbf{vt501}, respectively.
The detection accuracies in table are the best accuracies achieved by each detection method using values of ($t_{match}=1.0$), ($t_{classification}=1.0$), and ($d_{match}=2$).
Lastly, under each labeling scheme, the number of apps in each dataset (out of \textbf{Total Apps}), may differ, as seen in table \ref{tab:reference}.
The percentage under each detection accuracy depicts the percentage of apps each method managed to classify out of remaining apps in the dataset after applying the labeling scheme.}
\label{tab:demo}
\resizebox{\textwidth}{!}{
\begin{tabular}{|c|c|c|c|c||c|c|c|c||c|c|c|c||c|c|c|c||c|}
\hline
\rowcolor[HTML]{C0C0C0} 
\textbf{Detection Methods} & \multicolumn{4}{c||}{\cellcolor[HTML]{C0C0C0}\emph{Quick Matching}} & \multicolumn{4}{c||}{\cellcolor[HTML]{C0C0C0}\emph{Probabilistic Classification}} & \multicolumn{4}{c||}{\cellcolor[HTML]{C0C0C0}\emph{Deep Matching}} & \multicolumn{4}{c||}{\cellcolor[HTML]{C0C0C0}\emph{\dejavu\ Ensemble}} & \cellcolor[HTML]{C0C0C0} \\ \cline{1-17}
\rowcolor[HTML]{EFEFEF} 
\cellcolor[HTML]{C0C0C0}\textbf{Labeling Scheme ($\sigma_{label}$)} & \textbf{org} & \textbf{vt11} & \textbf{vt50} & \textbf{vt501} & \textbf{org} & \textbf{vt11} & \textbf{vt50} & \textbf{vt501} & \textbf{org} & \textbf{vt11} & \textbf{vt50} & \textbf{vt501} & \textbf{org} & \textbf{vt11} & \textbf{vt50} & \textbf{vt501} & \multirow{-2}{*}{\cellcolor[HTML]{C0C0C0}\textbf{Total Apps}} \\ \hline
 & 0.89 & \textbf{0.98} & 0.99 & 0.99 & 0.70 & \textbf{0.78} & 0.51 & 0.58 & 0.98 & \textbf{0.96} & 1.0 & 1.0 & 0.83 & \textbf{0.92} & 0.98 & 0.99 &  \\ \cline{2-17}
\multirow{-2}{*}{\begin{tabular}[c]{@{}c@{}}\emph{Piggybacked}\\ (confusion/malicious)\end{tabular}} & 71\% & \textbf{49\%} & 99\% & 87\% & 97\% & \textbf{80\%} & 55\% & 83\% & 4\% & \textbf{4\%} & 5\% & 0.7\% & 97\% & \textbf{91\%} & 96\% & 97\% & \multirow{-2}{*}{1399} \\ \hline
 & 0.99 & \textbf{0.95} & 0.85 & 0.87 & 0.37 & \textbf{0.45} & 0.56 & 0.27 & 0.64 & \textbf{0.91} & 1.0 & 0.91 & 0.97 & \textbf{0.90} & 0.69 & 0.93 &  \\ \cline{2-17}
\multirow{-2}{*}{\begin{tabular}[c]{@{}c@{}}\emph{Original}\\ (reference/benign)\end{tabular}} & 95\% & \textbf{67\%} & 21\% & 19\% & 46\% & \textbf{65\%} & 47\% & 72\% & 6\% & \textbf{7\%} & 7\% & 5\% & 98\% & \textbf{89\%} & 60\% & 98\% & \multirow{-2}{*}{1355} \\ \hline
 & 1.0 & \textbf{1.0} & 1.0 & 1.0 & 0.74 & \textbf{0.93} & 0.99 & 0.99 & 1.0 & \textbf{1.0} & 1.0 & 1.0 & 0.74 & \textbf{0.93} & 0.99 & 0.99 &  \\ \cline{2-17}
\multirow{-2}{*}{\begin{tabular}[c]{@{}c@{}}\emph{Malgenome}\\ (conventional/malicious)\end{tabular}} & 2\% & \textbf{2\%} & 2\% & 2\% & 99\% & \textbf{67\%} & 78\% & 93\% & 2\% & \textbf{1\%} & 1\% & 1\% & 99\% & \textbf{68\%} & 79\% & 93\% & \multirow{-2}{*}{1234} \\ \hline
\end{tabular}
}
\end{table*}

\begin{table*}[]
\caption{The detection accuracies scored by each \dejavu\ detection method along with the \dejavu\ Ensemble under different labeling schemes.
The numbers and percentages under the accuracies depict the number of apps all methods managed to classify. 
Under the \emph{vt50p-vt1} scheme (abbreviated \textbf{vt501}), the number of apps remaining in the dataset differs. 
So, the same number of apps classified under such a labeling scheme might yield a different percentage than the other schemes.}
\label{tab:summary}
\resizebox{\textwidth}{!}{
\begin{tabular}{|c|c|c|c|c||c|c|c|c||c|c|c|c||c|c|c|c||c|}
\hline
\rowcolor[HTML]{C0C0C0} 
\textbf{Detection Methods} & \multicolumn{4}{c||}{\cellcolor[HTML]{C0C0C0}\emph{Quick Matching}} & \multicolumn{4}{c||}{\cellcolor[HTML]{C0C0C0}\emph{Probabilistic Classification}} & \multicolumn{4}{c||}{\cellcolor[HTML]{C0C0C0}\emph{Deep Matching}} & \multicolumn{4}{c||}{\cellcolor[HTML]{C0C0C0}\emph{\dejavu\ Ensemble}} & \cellcolor[HTML]{C0C0C0} \\ \cline{1-17}
\rowcolor[HTML]{EFEFEF} 
\cellcolor[HTML]{C0C0C0}\textbf{Labeling Scheme ($\sigma_{label}$)} & \textbf{org} & \textbf{vt11} & \textbf{vt50} & \textbf{vt501} & \textbf{org} & \textbf{vt11} & \textbf{vt50} & \textbf{vt501} & \textbf{org} & \textbf{vt11} & \textbf{vt50} & \textbf{vt501} & \textbf{org} & \textbf{vt11} & \textbf{vt50} & \textbf{vt501} & \multirow{-2}{*}{\cellcolor[HTML]{C0C0C0}\textbf{Total Apps}} \\ \hline
 & 0.68 & \textbf{1.0} & 1.0 & 1.0 & 0.78 & \textbf{0.74} & 0.75 & 0.74 & \textcolor{red}{0.0} & \textcolor{red}{0.0} & \textcolor{red}{0.0} & \textcolor{red}{0.0} & 0.72 & \textbf{0.99} & 0.97 & 0.99 &  \\ \cline{2-17}
\multirow{-2}{*}{\begin{tabular}[c]{@{}c@{}}\emph{Piggybacked}\\ (confusion/malicious)\end{tabular}} & \multicolumn{3}{c|}{112 (8\%)} & 112 (38\%) & \multicolumn{3}{c|}{108 (8\%)} & 108 (36\%) & \multicolumn{4}{c||}{\textcolor{red}{0\%}} & \multicolumn{3}{c|}{267 (19\%)} & 267 (90\%) & \multirow{-2}{*}{1399} \\ \hline
 & 0.99 & \textbf{0.93} & 0.93 & 0.93 & 0.53 & \textbf{0.53} & 0.61 & 0.52 & 0.0 & 0.0 & 1.0 & 1.0 & 0.99 & \textbf{0.94} & 0.75 & 0.94 &  \\ \cline{2-17}
\multirow{-2}{*}{\begin{tabular}[c]{@{}c@{}}\emph{Original}\\ (reference/benign)\end{tabular}} & \multicolumn{3}{c|}{149 (11\%)} & 149 (17\%) & \multicolumn{3}{c|}{188 (14\%)} & 188 (22\%) & \multicolumn{3}{c|}{1 (0.07\%)} & 1 (0.1\%) & \multicolumn{3}{c|}{468 (34\%)} & 468 (54\%) & \multirow{-2}{*}{1355} \\ \hline
 & 1.0 & \textbf{1.0} & 1.0 & 1.0 & 0.91 & \textbf{0.91} & 0.99 & 0.99 & \textcolor{red}{0.0} & \textcolor{red}{0.0} & \textcolor{red}{0.0} & \textcolor{red}{0.0} & 0.90 & \textbf{0.93} & 0.99 & 0.99 &  \\ \cline{2-17}
\multirow{-2}{*}{\begin{tabular}[c]{@{}c@{}}\emph{Malgenome}\\ (conventional/malicious)\end{tabular}} & \multicolumn{4}{c||}{13 (1\%)} & \multicolumn{4}{c||}{513 (41\%)} & \multicolumn{4}{c||}{\textcolor{red}{0\%}} & \multicolumn{4}{c||}{685 (55\%)} & \multirow{-2}{*}{1234} \\ \hline
\end{tabular}
}
\end{table*}
\egroup

\subsection{Malware Detection Experiments}
\label{subsec:experiments}
As discussed in section \ref{sec:dejavu}, each \dejavu\ detection method expects different parameters, which can take different values. 
To answer \textbf{RQ2} and \textbf{RQ3} objectively, we should keep the value of each detection method's parameter constant across different datasets and labeling schemes. 
However, we do not know beforehand the impact of varying the values of each parameter on the detection accuracies of each detection method, which is the concern of \textbf{RQ1}. 
\subsubsection{Varying Parameters vs. Detection Accuracies}
We ran each of the 3 detection methods against 3 test datasets, using 4 different labeling schemes, and varying 3 to 4 values of the main parameters of each detection method (i.e., $t_{match} \in \{0.7, 0.8, 0.9, 1.0\}$ for \emph{quick matching}, $t_{classification}  \in \{0.7, 0.8, 0.9, 1.0\}$ for \emph{probabilistic classification}, and $d_{match} \in \{1, 2, 3\}$ for \emph{deep matching}). 
That yields a total of $3\times 4\times (2\times 4 + 3)=132$ experiments. 

The results of such experiments suggest that the detection methods \emph{quick matching}, \emph{probabilistic classification}, and \emph{deep matching} yield the best detection accuracies--regardless of the utilized dataset and labeling scheme--when the values of their respective parameters are ($t_{match}=1.0$), ($t_{classification}=1.0$), and ($d_{match}=2$). 
In summary, we noticed that the higher the value of a parameter, the more precise the detection method is. 
For example, increasing \emph{quick matching}'s ($t_{match}$) value helps the method increase its detection accuracy against the \emph{Original} dataset regardless of the adopted labeling scheme. 
We argue that such increase in the parameter's value instructs the method to harshen its criteria in matching test apps ($\alpha^{*}$) to apps in \dejavu's reference dataset ($\alpha$), effectively ruling out noisy apps and matching apps that are (almost) identical (i.e., with ($t_{match}\geq 0.9$). 
Similarly, higher values of ($d_{match}$) instruct \emph{deep matching} to not only match apps according to their metadata but to include the names of their files, components, classes, and methods as well, which intuitively yields more precise matchings and, in turn, classifications.
We plot such classification accuracies and summarize them in figure \ref{fig:line_acc} and table \ref{tab:parameters}, respectively, in appendix \ref{appendix:additional_figures}. 

Throughout this section, we plot and tabulate the detection accuracies achieved by each detection method using the aforementioned parameter values. 
Furthermore, we ran \dejavu's ensemble against the test datasets of \emph{Piggybacked}, \emph{Original}, and \emph{Malgenome} using such values. 

\subsubsection{Time vs. Detection Accuracies}
The data in table \ref{tab:demo} summarizes the detection accuracies achieved by \dejavu\ and its methods on different datasets under different labeling schemes. 
We use the data in this table to demonstrate the disarray varying the labeling scheme introduces to the composition of training datasets which ultimately affects the percentage of apps each detection method managed to analyze and the detection accuracies they yield. 
Without considering such percentages, it might seem that \emph{quick matching} performs better under the \emph{vt1-vt1} labeling scheme than on the \emph{original} one (i.e., 0.98 versus 0.89, respectively). 
However, under \emph{original}, \emph{quick matching} manages to classify 71\% of 1399 apps as opposed to 49\% under \emph{vt1-vt1}. 

To objectively compare the performance of detection methods, towards answering \textbf{RQ2} and \textbf{RQ3}, we considered the apps that each detection method managed to classify under all labeling schemes (i.e., an intersection), which we summarize in table \ref{tab:summary}.
A further demonstration of the impact of labeling schemes on the composition of training datasets and test results can be seen in the table. 
That is the number of apps all detection methods managed to classify depicts small percentages of the actual dataset. 
For example, only 112 apps (8\%) out of 1399 were classified by \emph{quick matching} under different labeling schemes. 
This phenomenon reaches an extreme for the \emph{deep matching} detection method which manages to classify completely different apps under each labeling scheme, yielding an intersection of \textbf{zero} apps.

Similar to table \ref{tab:reference}, we emboldened the labeling scheme that was adopted by the datasets' authors at the time of labeling and releasing the datasets to highlight the effect of time on the detection methods' accuracies.
In this case, the \emph{Piggybacked}, \emph{Original}, and \emph{Malgenome} datasets share the same scheme, viz.\ \emph{vt1-vt1}, or \textbf{vt11} for short.
We argue that time allows the \texttt{VirusTotal} scan results to mature or converge to a state that reflects each app's real intentions. 
That is to say, the older an app is, the longer the time analysts have to analyze it, and the more accurate its labels are expected to be, which should help different detection methods classify apps in a dataset more accurately. 

The detection accuracy scored by \emph{quick matching} on the \emph{Piggybacked} dataset under \emph{vt1-vt1} supports this hypothesis; the method's detection accuracy had an improvement of 0.32 (i.e., increase of 47\%), under the \emph{vt1-vt1} labeling scheme, albeit on 8\% of the dataset. 
As for \emph{Malgenome}, the performance of \emph{quick matching} remained the same. 
So, it seems the time positively affects the performance of \emph{quick matching} on malicious datasets. 
However, the exact opposite experience was encountered by the detection method on the \emph{Original} dataset. 
Under \emph{vt1-vt1}, as seen in table \ref{tab:reference}, the composition of the \emph{Original} dataset significantly differed, with only 63\% of the apps continued to be labeled as benign. 
To understand whether this structural modification affected the performance of \emph{quick matching}, we retrieved the \emph{Original} apps misclassified by the method under \emph{vt1-vt1}. 
We found that 100\% of such apps were compiled using \texttt{dexlib} and were identically matched to apps compiled using \texttt{dx} or \texttt{dexmerge}, yet the codebases of both apps differed. 
So, they were deemed malicious according to \dejavu's policy seen in line \texttt{8} of figure \ref{fig:overview}.
This scenario would manifest in case (a) the same legitimate developer of an app opts to update their codebase on the \texttt{smali} level and repackage their apps, or (b) more likely the \texttt{VirusTotal} scan results continue to label such apps as benign mistakenly. 
Under the \emph{original} labeling scheme, such apps would have been correctly classified as benign. 

The same performance exhibited by \emph{quick matching} was exhibited by \dejavu's ensemble. 
That is to say, on malicious datasets, time seems to enable \dejavu\ to score better accuracies, whereas the detection accuracies on the \emph{Original} dataset seems to worsen with time. 
We investigated the contribution of each individual detection method to detection accuracies scored by \dejavu\ as an ensemble (i.e., as seen in appendix \ref{appendix:additional_figures}), and found that \emph{quick matching} contributed the most to the correctly classified apps on the \emph{Piggybacked} and \emph{Original} datasets. 
This explains, we argue, why the performance of \emph{quick matching} was replicated on the ensemble method.
\subsubsection{Labeling Scheme vs. Detection Accuracies}
Towards answering \textbf{RQ3}, we investigated the impact of varying the labeling scheme on the detection accuracy of \dejavu's detection methods. 
Similar to the time dimension, we could not identify a unified pattern that explains the accuracies scored by all detection methods. 
In other words, one cannot assert, for example, that the performance of all detection methods improves after adopting the \emph{vt50p-vt1} labeling scheme. 
The only patterns we could identify were particular to a detection method, on a specific dataset, using certain labeling schemes. 
For instance, the accuracies of \emph{quick matching} and \dejavu's ensemble generally improve on the \emph{Piggybacked} and \emph{Malgenome} datasets the harsher the condition to deem an app as malicious gets. 
Apart from such individual patterns, the detection accuracies scored by different detection methods appear to be haphazard. 

We argue that varying the labeling scheme alters the labels in \dejavu's reference dataset, effectively yielding different versions of such datasets used to train the probabilistic classifier or match apps. 
For example, under the \emph{vt1-vt1} labeling scheme, the naive Bayes classifier used to classify apps in the \emph{Original} dataset is trained using 24552 malicious apps from the \emph{AMD} dataset, whereas the one under the \emph{vt50p-vt50p} scheme will be trained using 12765 malicious apps. 
Intuitively, those apps are expected to yield different accuracies. 

The mercurial performance of different detection methods under different labeling schemes on the same datasets is what we aspire to emboss in this paper. 
For example, consider the detection accuracies scored by \dejavu's ensemble on the \emph{Piggybacked} dataset. 
Under the \emph{original} labeling scheme, the method scores an accuracy of 0.72. 
However, the same method scores an accuracy of 0.99 on the exact same set of apps under the \emph{vt1-vt1} and \emph{vt50p-vt1}. 

So, depending on the adopted labeling scheme, the detection accuracy of the same detection method on the same dataset can differ by 37.5\% (i.e., from 0.72 to 0.99).  
This phenomenon, we believe, is of the utmost significance, especially since it implies that two independent groups of researchers might have--depending on the labeling perspectives they adopt--two completely different experiences with working on the same dataset using two similar or even identical methods. 
Not only does this encourages one of such hypothetical groups to mistakenly, yet justifiably, abandon their approach, it gives other researchers assessing the other group's work an incomplete picture of their detection method's capabilities.
\subsubsection{dejavu as an Ensemble}
With \textbf{RQ4}, we wish to investigate whether combining different detection methods as a sequential ensemble (i.e., as seen in figure \ref{fig:overview}), helps boost the detection accuracies. 
As discussed in the previous sections, \dejavu's ensemble detection method faced the same challenges encountered by the individual methods courtesy of varying the dimensions of time and labeling schemes. 
However, the ensemble method managed to outperform its individual counterparts in the number of apps classified. 
For all datasets, \dejavu's ensemble managed to classify more apps than individual detection methods. 

The reason behind this is that individual detection methods were designed to defer the classification of an app if they could not classify it with a confidence higher than a given value. 
Combining different methods increases the chances of classifying an app by one of the three methods.

\section{Discussion}
\label{sec:discussion}
In interpreting the results of our experiments, we could not find any patterns that link the detection accuracies scored by \dejavu's detection methods on different labeling schemes and datasets with one another. 
The lack of such patterns is exactly what we wish to demonstrate in this paper and its core contribution, viz.\ varying the freshness of labels (i.e., time), the labeling scheme, and the attack scenario (i.e., represented by the dataset used to test the method), makes the same detection method yield very different detection accuracies. 
So, to objectively assess the performance of a detection method, researchers should take into account those three dimensions, and evaluate their detection methods using the combinations of all values of such dimensions rather than focusing on the combination(s) that yield the best accuracies. 
As a demonstration, we applied this evaluation methodology on \dejavu\ and its detection methods and used the results of this evaluation to answer the research questions we postulated in section \ref{sec:evaluation}.


During our experiments, we noticed that, generally, varying the labeling scheme ($\sigma_{label}$) indeed affects the detection accuracies scored by \dejavu's detection methods. 
However, regardless of the adopted labeling scheme, we found that the higher the values of the parameters used by \dejavu's detection methods, the higher the detection accuracies they score, which is the concern of \textbf{RQ1}.
For example, the \emph{quick matching} detection method is guaranteed to yield the highest detection accuracies with values of ($t_{match}=1.0$). 
 
We argue that the higher the values of such parameters, the harsher the criteria the detection method applies to match a test app ($\alpha^{*}$) to one app or a class of reference apps ($\alpha$), and the more accurate the classification decision gets. 
For instance, a value of ($t_{classification}=1.0$) would instruct the \emph{probabilistic classification} detection method to return a label (i.e., malicious or benign), only if its classifier is 100\% confident in its decision, as opposed to 80\% or 70\% confidence for values of 0.8 or 0.7, respectively.

In table \ref{tab:reference}, we showed that time significantly alters the composition of different datasets. 
We expected such effect to be a reflection of the maturity of the scan results returned by \texttt{VirusTotal} \cite{mohaisen2014av}. 
That is, the older an Android app is, the more analysis it will be subject to, and the more accurate its labels will be. 
This should, in theory, help detection methods achieve better detection accuracies using labels drawn from more recent \texttt{VirusTotal} scans. 
Regarding (\textbf{RQ2}), the results of our experiments partly support this hypothesis.
We found that, by and large, \dejavu's detection methods perform better with time on \textbf{malicious} datasets (i.e., \emph{Piggybacked} and \emph{Malgenome}), especially \emph{quick matching} and the \emph{ensemble}. 

As for \textbf{RQ3}, similar to the time dimension, varying the labeling scheme ($\sigma_{label}$) had a noticeable effect on the detection accuracies of different detection methods. 
In this case, however, we could not identify patterns that explain the performances of all methods. 
Nonetheless, we demonstrate that by altering the labeling scheme, a given detection method can achieve on the exact same set of apps different detection accuracies with differences up to 0.32 in some cases. 
For example, as per table \ref{tab:summary}, the \dejavu\ ensemble detection method achieved on the \emph{Original} dataset a detection accuracy of 0.99 under the \emph{original} labeling scheme, 0.94 under \emph{vt1-vt1}, and 0.75 under \emph{vt50p-vt1}. 
So, using the same detection method and the same dataset of apps, researchers adopting a conservative labeling scheme of \emph{vt1-vt1} (i.e., an app is deemed malicious if only one \texttt{VirusTotal} scanner deems it so), those adopting a relaxed scheme of \emph{vt50p-vt50p}, and those relying on the \emph{original} labeling scheme adopted by the dataset authors will have very different experiences. 
We believe that such discrepancies stem from the way each detection method uses the reference dataset, whose composition is profoundly affected by varying the labeling scheme (i.e., as seen in table \ref{tab:reference}).
The aforementioned discrepancies, once again, emboss the effect of adopting different, labeling schemes on the performance of the same detection method on the same dataset. 
	
In answering \textbf{RQ4}, we found that certain detection methods outperform the others under different attack scenarios (i.e., as discussed in section \ref{subsec:attack_scenarios}). 
Using the results tabulated in table \ref{tab:summary}, we found that \emph{quick matching} performs better in most cases under the \emph{confusion} scenario. 
By definition, the \emph{confusion} scenario occurs whenever a repackaged app is being uploaded to a marketplace on which the original, benign app resides, in an attempt to confuse any vetting mechanism about the repackaged app's true intentions. 
So, based on this assumption, \emph{quick matching} attempts to match test apps ($\alpha^{*}$) to one, or more, of the benign apps in \dejavu's reference dataset ($\alpha$). 
Since the test app ($\alpha^{*}$) is designed to be similar to its benign counterpart ($\alpha$), \emph{quick matching} can effectively match apps and, using the techniques of codebase similarity and compiler fingerprinting, decide upon whether ($\alpha^{*}$) is a malicious version of ($\alpha$).

As for the \emph{probabilistic classification} method, we found that it performs poorly under the \emph{confusion} scenario. 
Nonetheless, we found that the method scores higher detection accuracies than the other detection methods under the \emph{conventional} scenario, under which the test app ($\alpha^{*}$) acts as an out-of-sample app that has never been seen before by a marketplace's app vetting mechanism.
Lastly, we found that the \emph{deep matching} detection method can be useful only upon being utilized as a complementary method (e.g., within the \dejavu\ ensemble). 
Despite scoring high detection accuracies (as seen in table \ref{tab:demo}), the method manages to analyze a small percentage of the datasets ranging from 1\% to 7\% when used individually. 

Within the malware analysis and detection domain, combining detection methods is usually expected to yield better results \cite{wang2018droidensemble,ledoux2015malware,spreitzenbarth2015mobile}. 
Consequently, we combined the individual detection methods of \emph{quick matching}, \emph{probabilistic classification}, and \emph{deep matching} as an ensemble, called \dejavu, and recorded its detection accuracies on different datasets and labeling schemes. 
To answer \textbf{RQ5}, we found that the ensemble method does not consistently out or underperform individual detection methods when considering accuracy as the performance metric. 
However, considering the larger number of apps it classifies, \dejavu's ensemble method can be considered to provide more comprehensive results.

Lastly, we postulated \textbf{RQ6} in an attempt to understand the nature of apps in the \emph{AMD} and \emph{Piggybacking} datasets that do not fit either criteria of the \emph{vt50p-vt1} labeling scheme. 
In other words, according to \texttt{VirusTotal} scan results, what are the apps that were deemed malicious by at least one scanner, yet were deemed malicious by less than 50\% of such scanners. 
For convenience, we refer to those apps as the \emph{gap apps}.
Our hypothesis was that such gap apps are so-called \emph{potentially unwanted software} or \emph{grayware} (e.g., \texttt{Adware}). 
Given that \texttt{VirusTotal} scanners seldom agree upon a common label, we used the data provided by Hurier et al.\ via their tool \emph{Euphony}\cite{hurier2017euphony} to get a consensus label for each gap app. 
Unfortunately, the data provided by \emph{Euphony} does not include consensus labels for all the identified gap apps.

\begin{table}[!t]
\centering
\caption{The number, percentage, and family distribution of apps in the \emph{Piggybacking} and \emph{AMD} datasets that fit neither criteria of the \emph{vt50p-vt1} labeling scheme.}
\label{tab:grayware}
\footnotesize
\begin{tabular}{|c|c|c|c|}
\hline
\rowcolor{gray!50}
\textbf{Dataset} & \textbf{Gap Apps} & \textbf{\emph{Euphony}} & \textbf{Top Families} \\ \hline
\emph{Piggybacking} & 1602 (58\%) & 906 & \begin{tabular}[c]{@{}c@{}}\textcolor{red}{\texttt{Adware} (68.87\%)}\\ \texttt{Trojan} (23.50\%)\\ \texttt{Spyware} (2.98\%)\end{tabular} \\ \hline
\emph{AMD} & 11788 (48\%) & 9250 & \begin{tabular}[c]{@{}c@{}}\textcolor{red}{\texttt{Adware} (95.36\%)}\\ \texttt{Trojan} (1.94\%)\\ \texttt{Monitor} (1.94\%)\end{tabular} \\ \hline
\end{tabular}
\end{table}

As seen in table \ref{tab:grayware}, applying the \emph{vt50p-vt1} labeling scheme reveals that about 58\% of the apps in the \emph{Piggybacking} dataset and about half of the apps in the \emph{AMD} dataset belong to this set of ambiguous apps. 
Using \emph{Euphony}, we managed to unify the labels of 906 and 9250 gap apps in the \emph{Piggybacking} and \emph{AMD} datasets, respectively. 
The results in the \textbf{Top Families} column reveal that the majority of such ambiguous gap apps is \texttt{Adware}. 
We postulate another research question: given the uncertainty of its nature, why is \texttt{Adware} considered malicious? 

Firstly, we argue that the malignancy of \texttt{Adware} is a matter of perspective. 
On the one hand, as long as the \texttt{Adware} payload does not disrupt the functionality of the app itself or overwhelm them with advertisements, users might not consider this breed of apps as malicious. 
On the other hand, developers that rely on revenues from advertisements embedded in their apps, yet have their apps repackaged to re-route such revenues to attackers will consider \texttt{Adware} as malicious. 

Secondly, the payloads of \texttt{Adware} are usually not malicious. 
That is they do not usually jeopardize the device's stability or leak any sensitive user information. 
Instead, they establish contact with remote URL's to display advertisements or download extra content. 
To decide upon whether such URL's are malicious, we extracted the URL's contacted by the gap apps in the \emph{AMD} and \emph{Piggybacking} datasets and used \texttt{VirusTotal} scans results of such URL's. 
Out of 553 URL's extracted from \emph{Piggybacking}'s gap apps, only 98 (17.72\%) were labeled as malicious by at least one \texttt{VirusTotal} scanner, with an average of 1.70 scanners out of 66.66 deeming the URL's malicious. 
The results for the \emph{AMD}'s gap apps were similar, viz.\ 1709 ($\approx$14\%) out of 12282 URL's were found malicious with an average of 1.27 positive scanners out of 66.61. 

We manually inspected a random sample of the URL's labeled as malicious in both datasets and found that the majority of such URL's pointed to advertisement engines and servers (e.g., \texttt{https://api.airpush.com}). 
Interestingly, some of the more suspicious URL's (e.g., \texttt{http://221.11.29.181} or \texttt{http://f5mv9t9x9wtx.pflexads.com}), were either labeled as benign or were not even scanned by any of the \texttt{VirusTotal} scanners.

\section{Threats to Validity}
\label{sec:threats}
We have identified four aspects that threaten the validity of our findings, which we discuss in the following paragraphs.

As a use case, we focus on Android repackaged malware due to the threat it poses to the Android ecosystem. 
Furthermore, given that Android repackaged malware can be encountered under different circumstances (as discussed in section \ref{subsec:attack_scenarios}), focusing on this breed of malware allows us to incorporate the dimension of attack scenario in our experiments. 
However, we argue that our evaluation framework is applicable to other types of Android malware and to other platforms (e.g., \texttt{Windows}-based malware), which we plan to investigate in future work.

Secondly, the sizes of the datasets we used in our evaluation (i.e., \emph{Piggybacked}, \emph{Original}, and \emph{Malgenome}), are relatively small. 
Those sizes further decrease upon focusing on the apps that were classified by a detection method under all different labeling schemes (i.e., table \ref{tab:summary}).
However, these are state-of-the-art datasets for the use case we have in this paper, and we consider it out of scope to increase the sizes of these datasets. Other use cases (e.g., \texttt{Windows}-based malware) may have larger datasets. Nevertheless, this is an important aspect to consider when using our evaluation methodology.

Thirdly, the detection methods and techniques utilized by \dejavu\ depict only a subset of all the methods devised to detect Android (repackaged) malware. 
For instance, we did not consider dynamic-based methods that consider the runtime behaviors of test apps. 
Moreover, we only used one classifier (i.e., naive Bayes), instead of an ensemble of classifiers, for example.
However, the objective of this paper is neither to compare all possible detection methods for Android (repackaged) malware nor to promote \dejavu\ as a particularly promising detection method. 
The core contribution of our work is to indicate that--regardless of the type, quality, or sophistication of the utilized detection method--varying the freshness of \texttt{VirusTotal}'s scan results, the scheme adopted to interpret them to label apps, and the dataset used to emulate a particular attack scenario significantly impacts the detection accuracy a detection method. 

Lastly, the 3 dimensions in our evaluation methodology are not to be considered an exhaustive list. 
We plan to research other aspects that might affect the performance of malware detection methods and incorporate them in our framework. 



\section{Related Work}
\label{sec:related}
In this section, we enumerate the research efforts we found are most relevant to ours.

\paragraph{\textbf{(Android) Malware Labeling}} 
There is a number of efforts that discuss the challenges facing the (Android) malware detection community due to the inconsistencies of labeling schemes and the volatility of data acquired from antiviral software/platforms. 
For example, in \cite{hurier2016lack}, Hurier et al.\ discuss the difficulty of acquiring a consensus vis-\`{a}-vis the label of a malicious Android app across different antiviral scanners. 
Similarly, Sebastian et al.\ in \cite{sebastian2016avclass} introduce a cross-platform approach to rank (Android) malware labels and remove any noisy or redundant aliases associated with a malware instance. 
A closer work to ours, despite not focusing on Android malware, is that of Mohaisen et al.\ in \cite{mohaisen2014av}. 
In this work, the labels returned by different antiviral scanners are assessed to unveil the danger of relying on incomplete, inconsistent, and incorrect
malware labels. 

We have noticed that different researchers are aware of the problems of using (a) inconsistent and outdated information acquired from platforms such as \texttt{VirusTotal}, and (b) the presence of different interpretations of such information to label Android apps \cite{li2018rebooting,wang2018beyond,wei2017deep,hurier2017euphony}. 
Nonetheless, none of the aforementioned studies, and, to the best of our knowledge, no studies investigate the impact of using outdated labels or different labeling schemes on the performance of a detection method even on the same dataset. 

	

\paragraph{\textbf{Attack Scenarios}}
In our experiments, we considered two scenarios under which Android repackaged malware can be encountered by detection or app vetting mechanisms, namely, \emph{conventional} scenario and \emph{confusion} scenario.
In the latter scenario, we assumed that attackers may elect to target marketplaces on which the original, benign versions of their malicious, repackaged instances reside to confuse any app vetting mechanisms using such benign apps as references of benign behaviors \cite{chen2018android,demontis2017yes,yang2017malware,grosse2017adversarial}. 
Numerous efforts focus on devising and demonstrating methods to repackage Android apps in a manner that evades detection by any detection methods. 
By and large, such efforts focus on evading detection by machine learning classifiers and, thus, assumes that attackers possess substantial knowledge about (a) the structure of the classifier they are up against (e.g., the model it uses, its parameters, and the features used to train it), and (b) the field of adversarial machine learning.

In section \ref{subsec:attack_scenarios}, we defined a more realistic attacker that does not possess that much information about the detection methods they are trying to evade. 
Consequently, the closest work to ours was that of Salem et al.\ in \cite{salem2018poking}. 
In their work, they demonstrated the effect of the \emph{confusion} scenario on the performance of an ensemble of machine learning classifiers that otherwise perform well (i.e., under the \emph{conventional} scenario).
They demonstrated that by testing an ensemble of classifiers that have been trained using apps from the \emph{Original} segment of the \emph{Piggybacking} dataset using apps from the \emph{Piggybacked} segment. 
Despite solely using the original labeling scheme of the dataset, their experiments show the impact of the \emph{confusion} scenario, which inspired us to evaluate \dejavu\ under similar circumstances.

\section{Conclusion}
\label{sec:conclusion}
In evaluating (Android) malware detection methods, researchers usually rely on (outdated) scan results obtained from \texttt{VirusTotal} to label the apps in their training and test datasets \cite{wei2017deep,li2017understanding,zhou2012dissecting}. 
Unfortunately, due to the lack of concrete standards or common practices that guide this labeling process, researchers rely on their subjective intuition to label apps. 
Using outdated labeling information and adopting different schemes to label apps, we argue, significantly affects the compositions of the datasets utilized by a malware detection method during evaluation, which leads to very different outcomes. 
Unfortunately, this phenomenon renders the effectiveness of a detection method as a matter of perspective. 
That is, depending on which side of this labeling kaleidoscope we are standing, the performance of a given detection method even on the same dataset might differ. 

To demonstrate volatility of the data acquired from \texttt{VirusTotal} and the effect of interpreting it differently on the performance of Android repackaged malware detection methods, we implemented a representative ensemble of three detection methods, called \dejavu, and used it to conduct more than 130 experiments using four different labeling schemes and three different datasets containing over 30,000 Android apps. 
The results obtained from our experiments indeed show that the detection accuracy achieved by one detection method might significantly vary depending on the freshness of the labels (e.g., acquired from \texttt{VirusTotal}), the labeling scheme, and the attack scenario, with differences up to 47\% (i.e., from 0.68 to 1.0). 

With our results and insights, we aspire to instigate the adoption of a methodology to evaluate the effectiveness of malware detection methods, that takes into consideration the freshness of labels, the different labeling schemes that can be adopted to label apps, and the different scenarios under which malware can be found, if applicable. 
In other words, we encourage researchers to adopt the following practices in evaluating the detection methods they devise.
Firstly, malware analysis and detection researchers need to take the time dimension and the evolution of \texttt{VirusTotal} scan results into consideration upon labeling their apps and re-analyze those apps or acquire the latest scan results prior to evaluating their methods. 
Secondly, to accommodate different labeling schemes and ensure the comprehensiveness of their methods, researchers are encouraged to document the effect of adopting different labeling schemes on the performance of their methods.
Needless to say, the more consistent the performance of a method across different labeling schemes, the steadier and more reliable it is. 
Lastly, researchers are encouraged to utilize different datasets of apps and scenarios (as seen in section \ref{subsec:attack_scenarios}), to ensure that their detection methods are resilient under different circumstances.
With that in mind, our primary focus for future work is to devise a measure that gives an overall assessment of the effectiveness of a detection method given its performance across different combinations of the dimensions of time, labeling scheme, and attack scenario.


{\normalsize \bibliographystyle{acm}
\bibliography{dejavu_usenix}}
\appendix
\section*{Appendices}
\addcontentsline{toc}{section}{Appendices}
\renewcommand{\thesubsection}{\Alph{subsection}}
\subsection{Static Features}
\label{appendix:static_features}
As discussed in section \ref{subsec:probabilistic_classification}, we trained our probabilistic classifiers using numerical features statically extracted from the APK archives of apps in the \emph{Piggybacking} (i.e, \emph{Piggybacked}+\emph{Original}), \emph{AMD}, \emph{Malgenome}, and \emph{GPlay} datasets. 
In total, each app was represented by a vector of 40 features, which are divided into four categories, namely, basic features, permission-based features, API call features, and miscellaneous features. 
The following list mimics the order of every feature in the feature vector.
\begin{enumerate}
	\item[] Basic features:
		\begin{enumerate}
			\item[1.] Minimum SDK version (supported by the app).
			\item[2.] Maximum SDK version (supported by the app).
			\item[3.] Total number of activities.
			\item[4.] Total number of services.
			\item[5.] Total number of broadcast receivers.
			\item[6.] Total number of content providers.
		\end{enumerate}
	\item[] Permission-based features:
		\begin{enumerate}
			\item[7.] Total request permissions.
			\item[8.] Ratio of Android permissions to total permissions.
			\item[9.] Ratio of Custom permissions to total permissions.
			\item[10.] Ratio of dangerous permissions to total permissions.
		\end{enumerate}
	\item[] API call features:
		\begin{enumerate}
			\item[11.] Total number of classes.
			\item[12.] Total number of methods.
			\item[13.] Count of calls to methods in the \texttt{android.accounts.AccountManager} package.
			\item[14.] Count $\dots$ \texttt{android.app.Activity} package.
			\item[15.] Count $\dots$ \texttt{android.app.DownloadManager} package.
			\item[16.] Count $\dots$ \texttt{android.app.IntentService} package.
			\item[17.] Count $\dots$ \texttt{android.content.ContentResolver} package.
			\item[18.] Count $\dots$ \texttt{android.content.ContextWrapper} package.
			\item[19.] Count $\dots$ \texttt{android.content.pm.PackageInstaller} package.
			\item[20.] Count $\dots$ \texttt{android.database.sqlite.} \texttt{SQLiteDatabase} package.
			\item[21.] Count $\dots$ \texttt{android.hardware.Camera} package.
			\item[22.] Count $\dots$ \texttt{android.hardware.display.} \texttt{DisplayManager} package.
			\item[23.] Count $\dots$ \texttt{android.location.Location} package.
			\item[24.] Count $\dots$ \texttt{android.media.AudioRecord} package.
			\item[25.] Count $\dots$ \texttt{android.media.MediaRecorder} package.
			\item[26.] Count $\dots$ \texttt{android.net.Network} package.
			\item[27.] Count $\dots$ \texttt{android.net.NetworkInfo} package.
			\item[28.] Count $\dots$ \texttt{android.net.wifi.WifiInfo} package.
			\item[29.] Count $\dots$ \texttt{android.net.wifi.WifiManager} package.
			\item[30.] Count $\dots$ \texttt{android.os.PowerManager} package.
			\item[31.] Count $\dots$ \texttt{android.os.Process} package.
			\item[32.] Count $\dots$ \texttt{android.telephony.SmsManager} package.
			\item[33.] Count $\dots$ \texttt{android.widget.Toast} package.
			\item[34.] Count $\dots$ \texttt{dalvik.system.DexClassLoader} package.
			\item[35.] Count $\dots$ \texttt{dalvik.system.PathClassLoader} package.
			\item[36.] Count $\dots$ \texttt{java.lang.class} package.
			\item[37.] Count $\dots$ \texttt{java.lang.reflect.Method} package.
			\item[38.] Count $\dots$ \texttt{java.net.HttpCookie} package.
			\item[39.] Count $\dots$ \texttt{java.net.URL.openConnection} package.
		\end{enumerate}
	\item[] Miscellaneous features:
		\begin{enumerate}
			\item[40.] Zero-based index of the compiler used to compile the app from (\texttt{dx}, \texttt{dexmerge}, \texttt{dexlib 1.x}, \texttt{dexlib 2.x}, \texttt{Jack 4.x}, or unknown).
		\end{enumerate}
\end{enumerate}

\begin{table*}[]
\caption{A summary of the values of ($t_{match}$), ($t_{classification}$), and ($d_{match}$) that respectively helped the \emph{quick matching}, \emph{probabilistic classification}, and \emph{deep matching} detection methods yield the highest accuracies on different datasets and labeling schemes.}
\label{tab:parameters}
\small
\resizebox{\textwidth}{!}{
\begin{tabular}{|c|c|c|c|c|c|c|c|c|c|}
\hline
\rowcolor[HTML]{C0C0C0} 
\textbf{Detection Methods} & \multicolumn{3}{c|}{\cellcolor[HTML]{C0C0C0}\begin{tabular}[c]{@{}c@{}}\emph{Quick Matching}\\ ($t_{match}$)\end{tabular}} & \multicolumn{3}{c|}{\cellcolor[HTML]{C0C0C0}\begin{tabular}[c]{@{}c@{}}\emph{Probabilistic Classification}\\ ($t_{classification}$)\end{tabular}} & \multicolumn{3}{c|}{\cellcolor[HTML]{C0C0C0}\begin{tabular}[c]{@{}c@{}}\emph{Deep Matching}\\ ($d_{match}$)\end{tabular}} \\ \hline
\rowcolor[HTML]{EFEFEF} 
\cellcolor[HTML]{C0C0C0}\textbf{Labeling Scheme ($\sigma_{label}$)} & \emph{vt1-vt1} & \emph{vt50p-vt50p} & \emph{vt50p-vt1} & \emph{vt1-vt1} & \emph{vt50p-vt50p} & \emph{vt50p-vt1} & \emph{vt1-vt1} & \emph{vt50p-vt50p} & \emph{vt50p-vt1} \\ \hline
\begin{tabular}[c]{@{}c@{}}\emph{Piggybacked}\\ (confusion/malicious)\end{tabular} & $\geq$ 0.8 & $\geq$ 0.9 & $\geq$ 0.7 & 1.0 & 1.0 & 1.0 & 3 & 2 & 2 \\ \hline
\begin{tabular}[c]{@{}c@{}}\emph{Original}\\ (reference/benign)\end{tabular} & $\geq$ 0.7 & 1.0 & $\geq$ 0.7 & 1.0 & 1.0 & 1.0 & 2 & 2 & 3 \\ \hline
\begin{tabular}[c]{@{}c@{}}\emph{Malgenome}\\ (conventional/malicious)\end{tabular} & $\geq$ 0.7 & $\geq$ 0.7 & $\geq$ 0.8 & 1.0 & 1.0 & 1.0 & 2 & 2 & 2 \\ \hline
\end{tabular}
}
\end{table*}

\begin{table*}[]
\caption{This table shows the contribution of each individual method (e.g., \emph{quick matching}), to the overall detection accuracy of \dejavu's ensemble. 
That is, how many of the correctly classified apps in each dataset-labeling scheme combination is attributed to each detection method, and how long (on average) did it take each method to correctly classify an app.}
\label{tab:pies}
\small
\resizebox{\textwidth}{!}{
\begin{tabular}{|c|c|c|c|c|c|c|c|c|c|}
\hline
\rowcolor[HTML]{C0C0C0} 
\textbf{Detection Methods} & \multicolumn{3}{c|}{\cellcolor[HTML]{C0C0C0}\begin{tabular}[c]{@{}c@{}}\emph{Quick Matching}\\ (seconds)\end{tabular}} & \multicolumn{3}{c|}{\cellcolor[HTML]{C0C0C0}\begin{tabular}[c]{@{}c@{}}\emph{Probabilistic Classification}\\ (seconds)\end{tabular}} & \multicolumn{3}{c|}{\cellcolor[HTML]{C0C0C0}\begin{tabular}[c]{@{}c@{}}\emph{Deep Matching}\\ (seconds)\end{tabular}} \\ \hline
\rowcolor[HTML]{EFEFEF} 
\cellcolor[HTML]{C0C0C0}\textbf{Labeling Scheme ($\sigma_{label}$)} & \emph{vt1-vt1} & \emph{vt50p-vt50p} & \emph{vt50p-vt1} & \emph{vt1-vt1} & \emph{vt50p-vt50p} & \emph{vt50p-vt1} & \emph{vt1-vt1} & \emph{vt50p-vt50p} & \emph{vt50p-vt1} \\ \hline
\begin{tabular}[c]{@{}c@{}}\emph{Piggybacked}\\ (confusion/malicious)\end{tabular} & \begin{tabular}[c]{@{}c@{}}63\%\\ (19.89)\end{tabular} & \begin{tabular}[c]{@{}c@{}}96\%\\ (8.81)\end{tabular} & \begin{tabular}[c]{@{}c@{}}81\%\\ (7.28)\end{tabular} & \begin{tabular}[c]{@{}c@{}}37\%\\ (15.36)\end{tabular} & \begin{tabular}[c]{@{}c@{}}3\%\\ (8.99)\end{tabular} & \begin{tabular}[c]{@{}c@{}}18\%\\ (5.82)\end{tabular} & \begin{tabular}[c]{@{}c@{}}0.1\%\\ (317.2)\end{tabular} & \begin{tabular}[c]{@{}c@{}}0.1\%\\ (64.8)\end{tabular} & \begin{tabular}[c]{@{}c@{}}0.3\%\\ (194.0)\end{tabular} \\ \hline
\begin{tabular}[c]{@{}c@{}}\emph{Original}\\ (reference/benign)\end{tabular} & \begin{tabular}[c]{@{}c@{}}75\%\\ (5.78)\end{tabular} & \begin{tabular}[c]{@{}c@{}}42\%\\ (5.22)\end{tabular} & \begin{tabular}[c]{@{}c@{}}98\%\\ (16.4)\end{tabular} & \begin{tabular}[c]{@{}c@{}}24\%\\ (7.53)\end{tabular} & \begin{tabular}[c]{@{}c@{}}52\%\\ (12.18)\end{tabular} & \begin{tabular}[c]{@{}c@{}}1\%\\ (75.61)\end{tabular} & \begin{tabular}[c]{@{}c@{}}0.7\%\\ (125.5)\end{tabular} & \begin{tabular}[c]{@{}c@{}}5\%\\ (100.2)\end{tabular} & \begin{tabular}[c]{@{}c@{}}0.2\%\\ (160.7)\end{tabular} \\ \hline
\begin{tabular}[c]{@{}c@{}}\emph{Malgenome}\\ (conventional/malicious)\end{tabular} & \begin{tabular}[c]{@{}c@{}}2\%\\ (2.66)\end{tabular} & \begin{tabular}[c]{@{}c@{}}4\%\\ (3.14)\end{tabular} & \begin{tabular}[c]{@{}c@{}}2\%\\ (3.1)\end{tabular} & \begin{tabular}[c]{@{}c@{}}97\%\\ (1.49)\end{tabular} & \begin{tabular}[c]{@{}c@{}}95\%\\ (2.87)\end{tabular} & \begin{tabular}[c]{@{}c@{}}98\%\\ (1.93)\end{tabular} & \begin{tabular}[c]{@{}c@{}}0.6\%\\ (89.4)\end{tabular} & \begin{tabular}[c]{@{}c@{}}0.5\%\\ (83.7)\end{tabular} & \begin{tabular}[c]{@{}c@{}}0\%\\ (0.0)\end{tabular} \\ \hline
\end{tabular}
}
\end{table*}

\subsection{Additional Figures}
\label{appendix:additional_figures}
\subsubsection{dejavu Ensemble Accuracies and Parameters}
\begin{sidewaysfigure*}
\centering
\begin{subfigure}{0.32\textwidth}
	\includegraphics[scale=0.45]{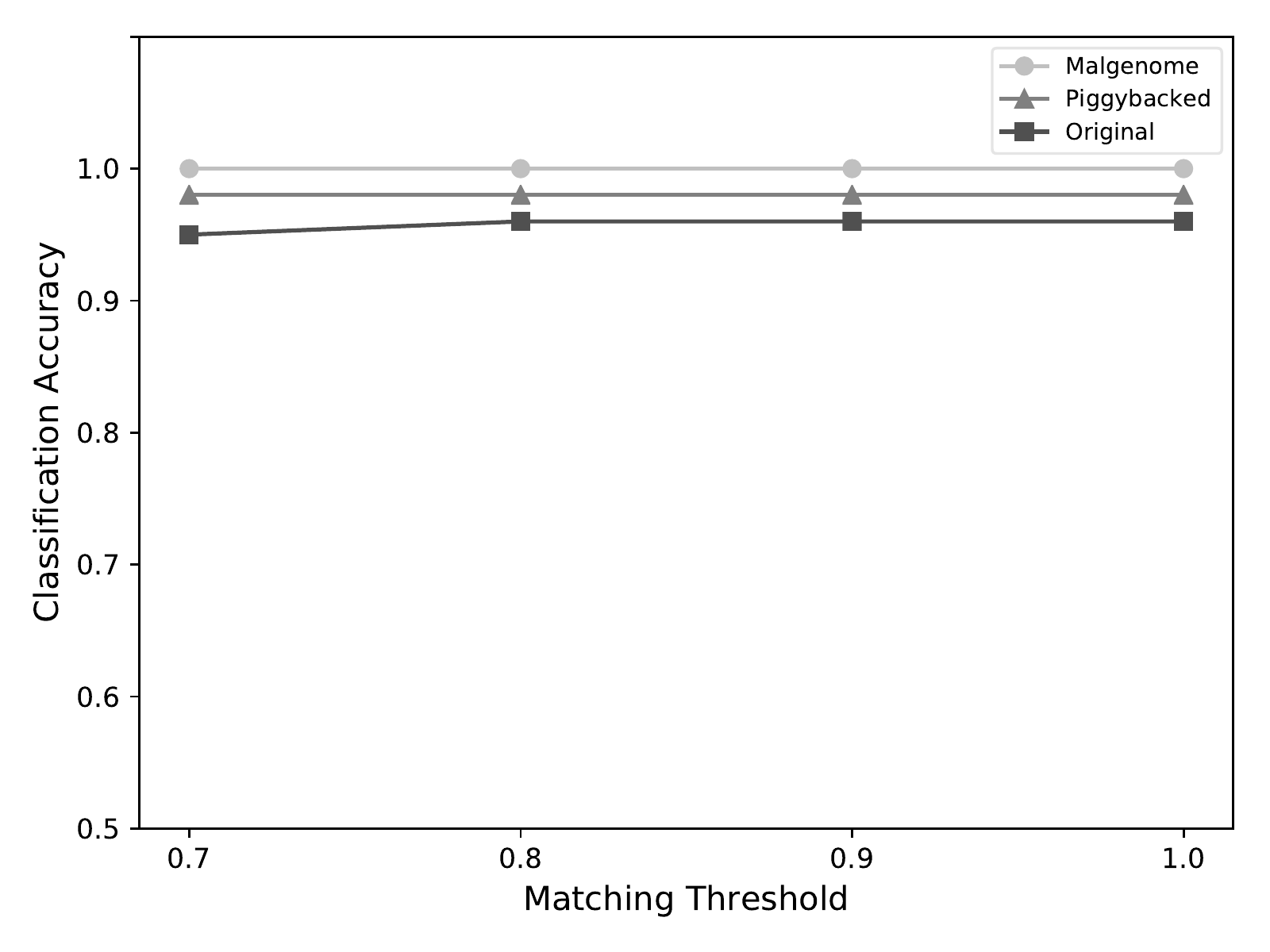}
	\caption{\emph{quick matching} with \emph{vt1-vt1}}
	\label{fig:qm_vt1vt1}
\end{subfigure}
\begin{subfigure}{0.32\textwidth}
	\includegraphics[scale=0.45]{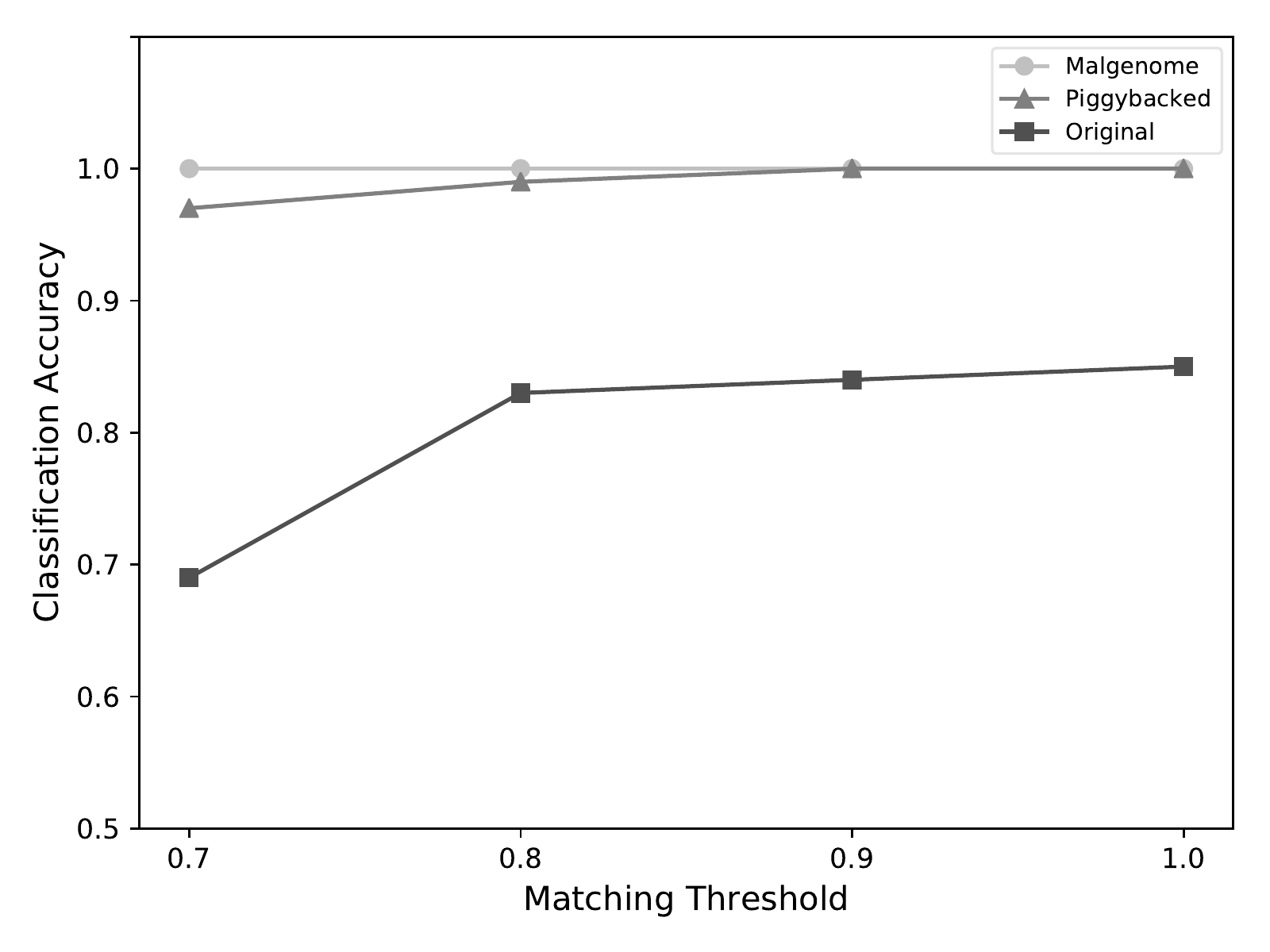}
	\caption{\emph{quick matching} with \emph{vt50p-vt50p}}
	\label{fig:qm_vt50pvt50p}
\end{subfigure}
\begin{subfigure}{0.32\textwidth}
	\includegraphics[scale=0.45]{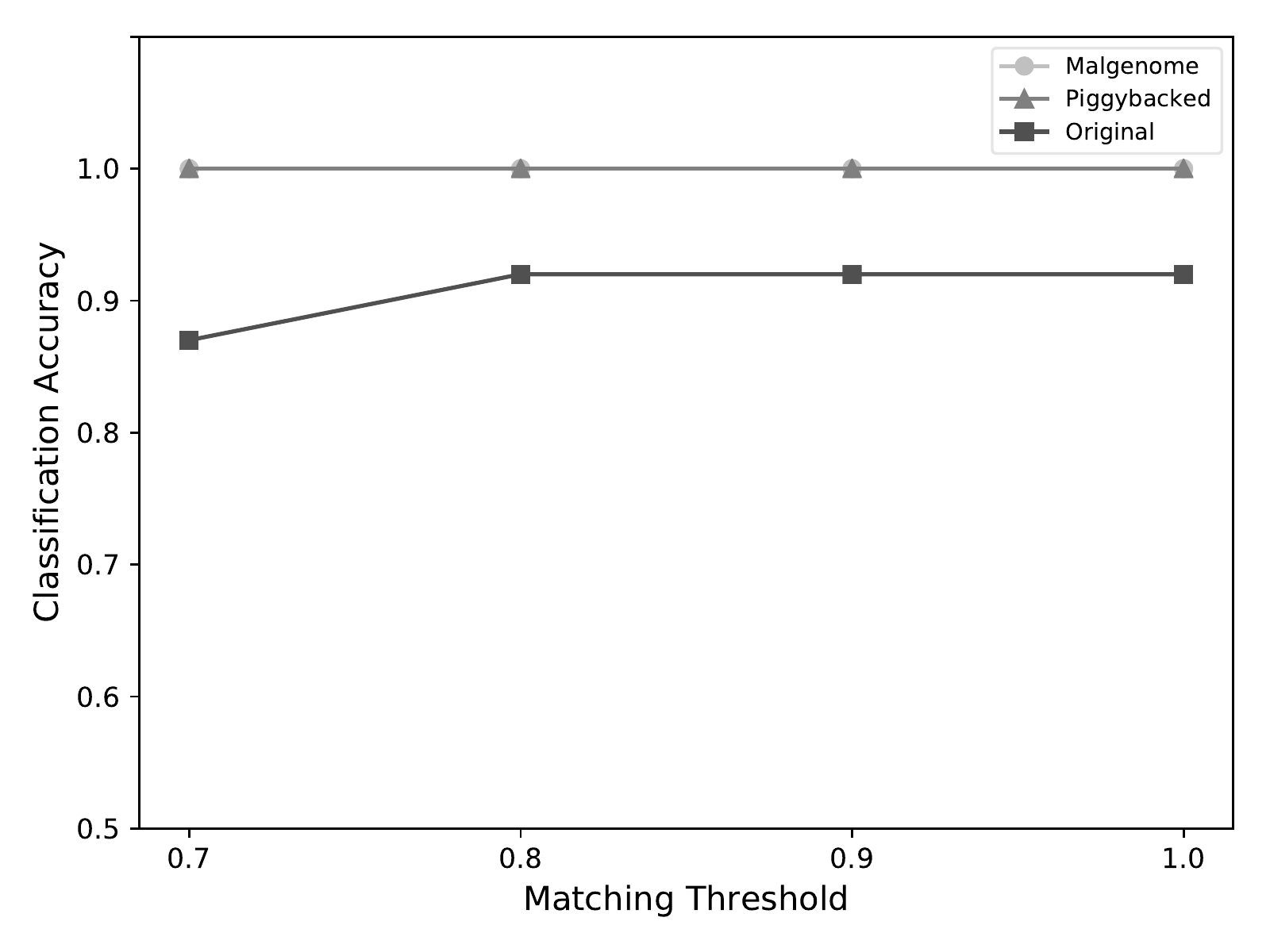}
	\caption{\emph{quick matching} with \emph{vt50p-vt1}}
	\label{fig:qm_vt50pvt1}
\end{subfigure}
\begin{subfigure}{0.32\textwidth}
	\includegraphics[scale=0.45]{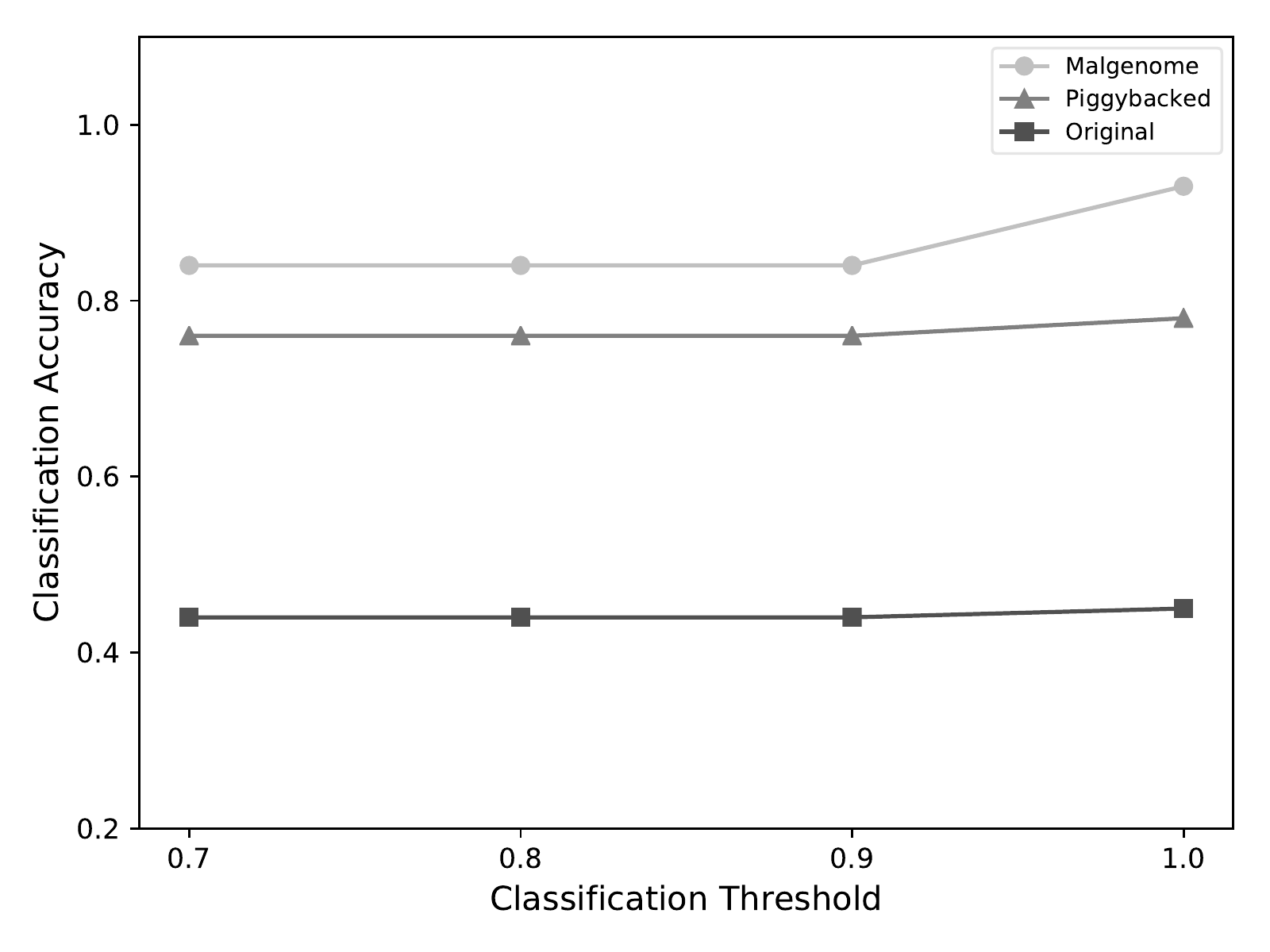}
	\caption{\emph{probabilistic classification} with \emph{vt1-vt1}}
	\label{fig:pc_vt1vt1}
\end{subfigure}
\begin{subfigure}{0.32\textwidth}
	\includegraphics[scale=0.45]{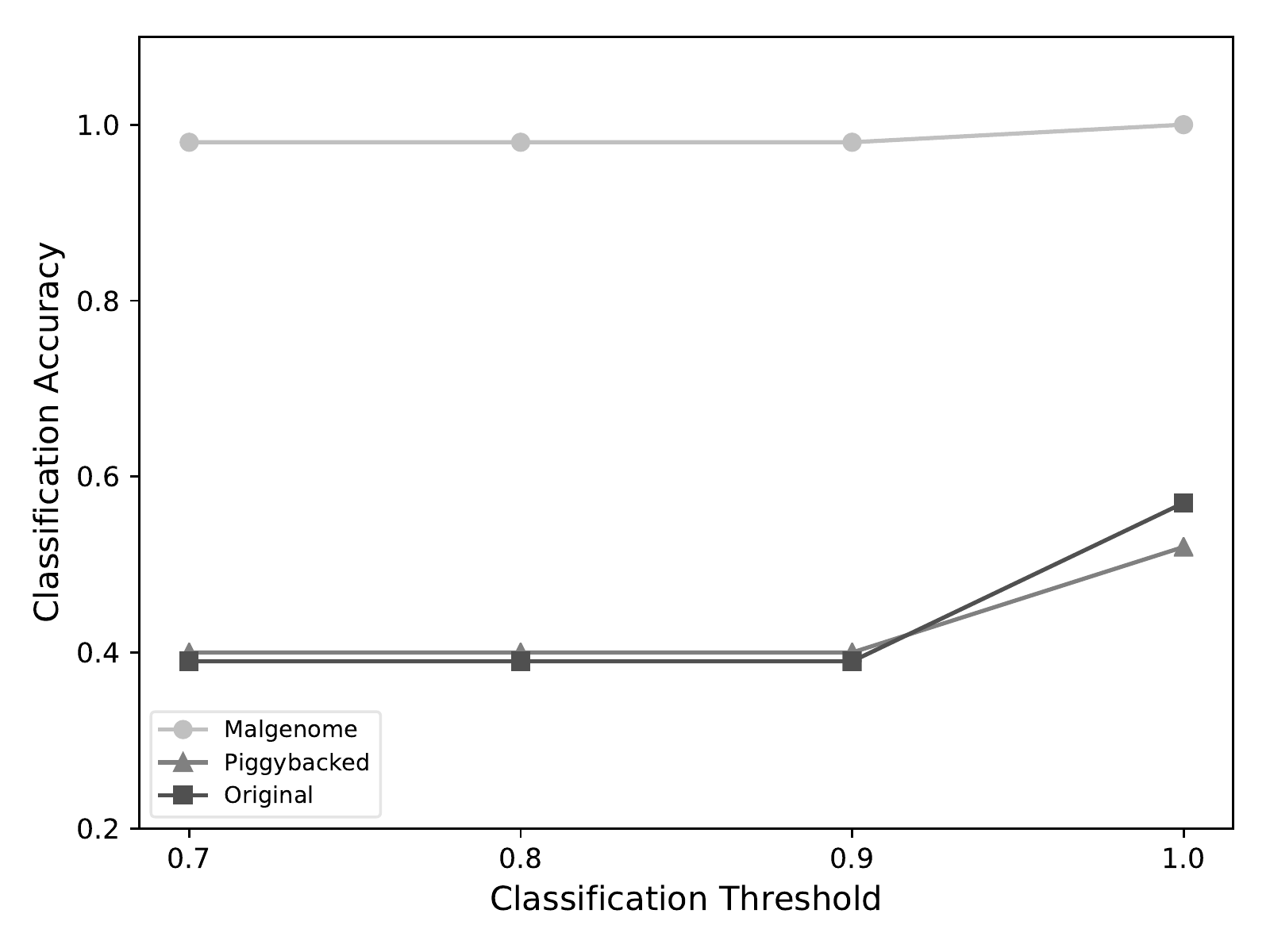}
	\caption{\emph{probabilistic classification} with \emph{vt50p-vt50p}}
	\label{fig:pc_vt50pvt50p}
\end{subfigure}
\begin{subfigure}{0.32\textwidth}
	\includegraphics[scale=0.45]{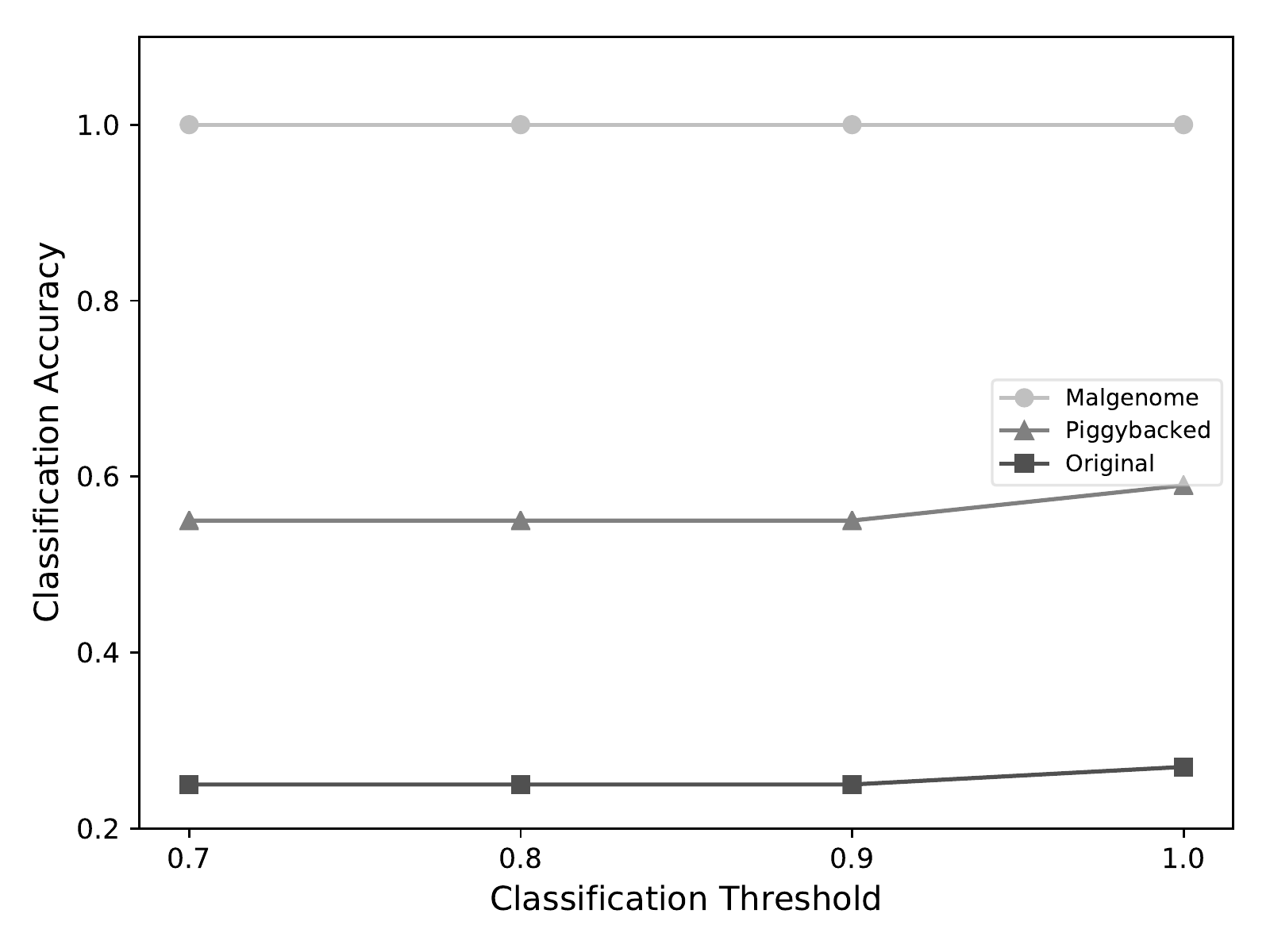}
	\caption{\emph{probabilistic classification} with \emph{vt50p-vt1}}
	\label{fig:pc_vt50pvt1}
\end{subfigure}
\begin{subfigure}{0.32\textwidth}
	\includegraphics[scale=0.45]{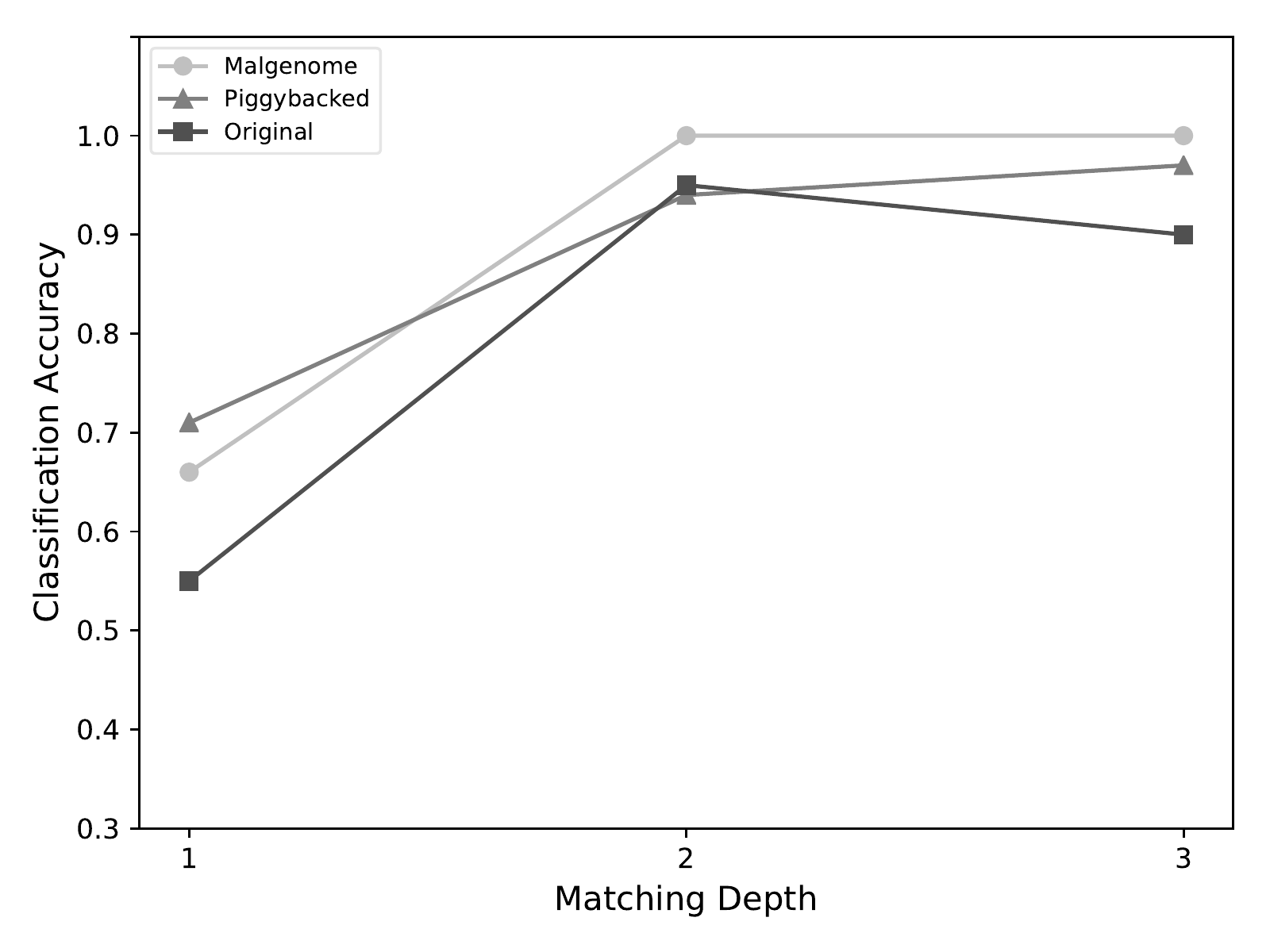}
	\caption{\emph{deep matching} with \emph{vt1-vt1}}
	\label{fig:dm_vt1vt1}
\end{subfigure}
\begin{subfigure}{0.32\textwidth}
	\includegraphics[scale=0.45]{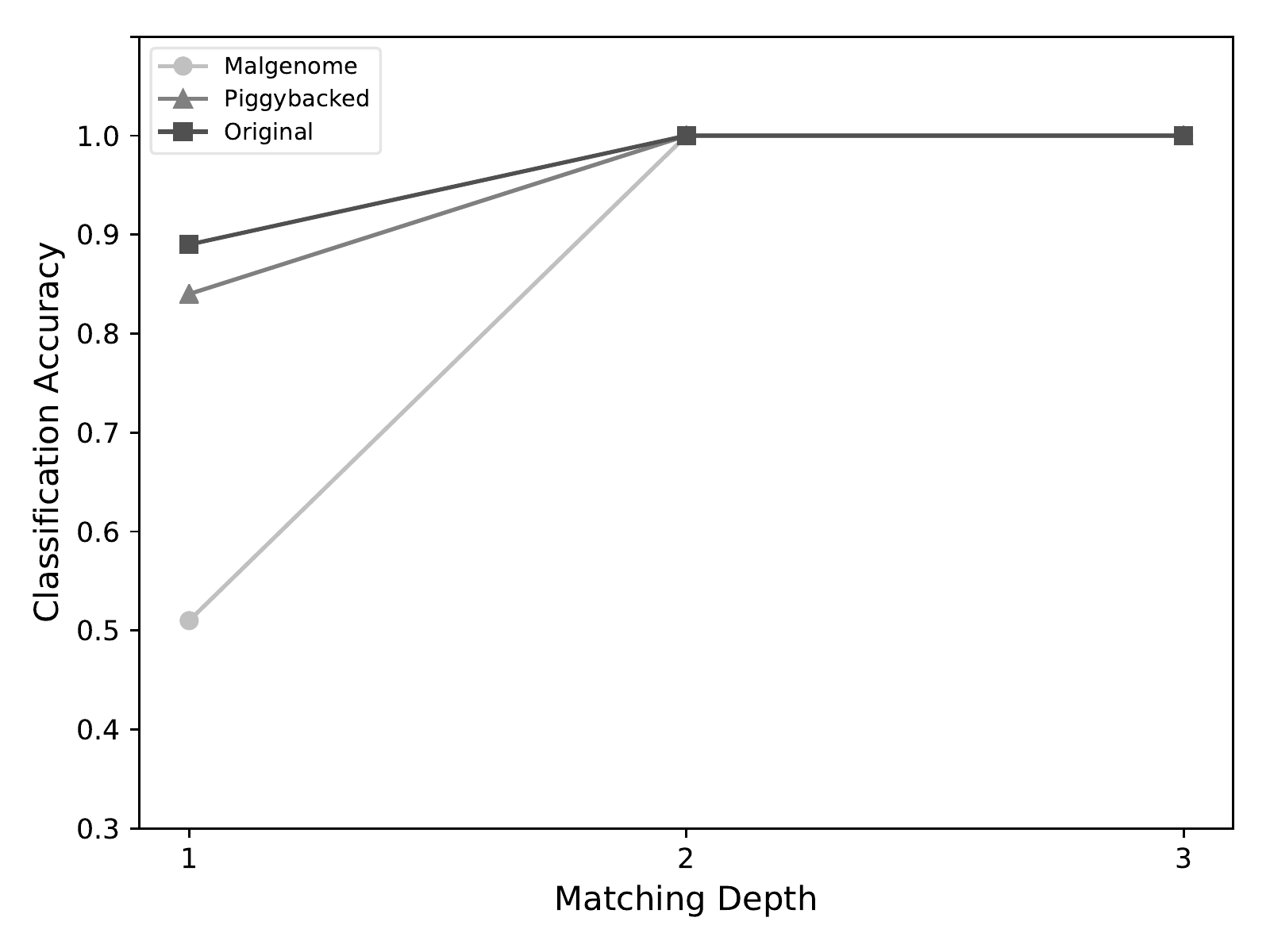}
	\caption{\emph{deep matching} with \emph{vt50p-vt50p}}
	\label{fig:dm_vt50pvt50p}
\end{subfigure}
\begin{subfigure}{0.32\textwidth}
	\includegraphics[scale=0.45]{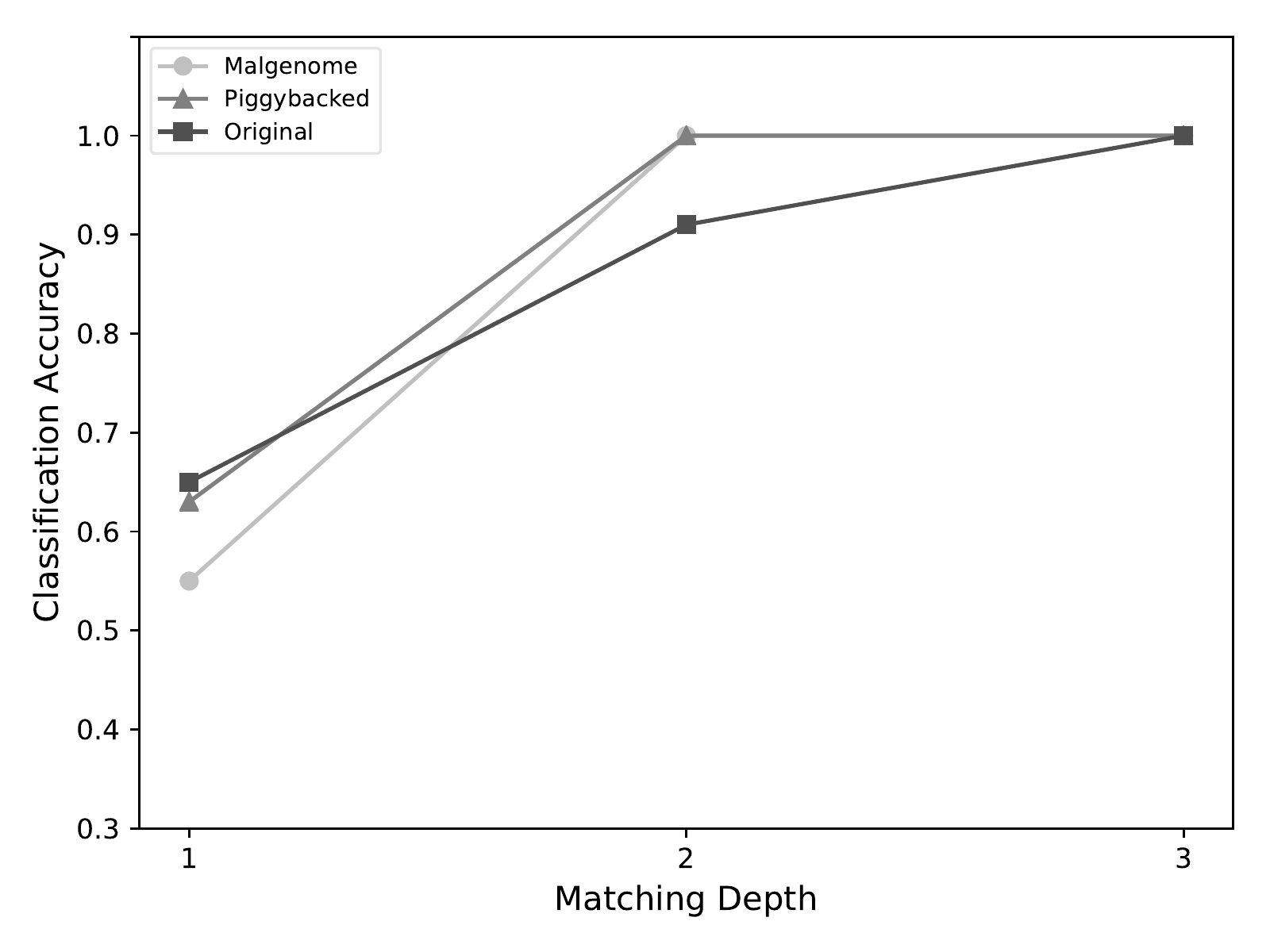}
	\caption{\emph{deep matching} with \emph{vt50p-vt1}}
	\label{fig:dm_vt50pvt1}
\end{subfigure}
\caption{The detection accuracies of each individual \dejavu\ detection method on each dataset and labeling scheme.}
\label{fig:line_acc}
\end{sidewaysfigure*}

To address research question (\textbf{RQ1}), we conducted a series of experiments to investigate any relationship between different labelings scheme and the parameters adopted by \dejavu's individual detection methods (i.e., ($t_{match}$), ($t_{classification}$), and ($d_{match}$)), and whether this relationship affects the detection accuracies scored by each method on the \emph{Piggybacked}, \emph{Original}, and \emph{Malgenome} datasets. 
In figure \ref{fig:line_acc}, we plot this relationship and use it to extract the parameter values that helped each detection method achieve the highest detection accuracy on each dataset-labeling scheme combination. 
Furthermore, we tabulate such values in table \ref{tab:parameters}. 

\subsubsection{dejavu Detection Method Decomposition}
\begin{sidewaysfigure*}
\centering
\begin{subfigure}{0.32\textwidth}
	\includegraphics[scale=0.45]{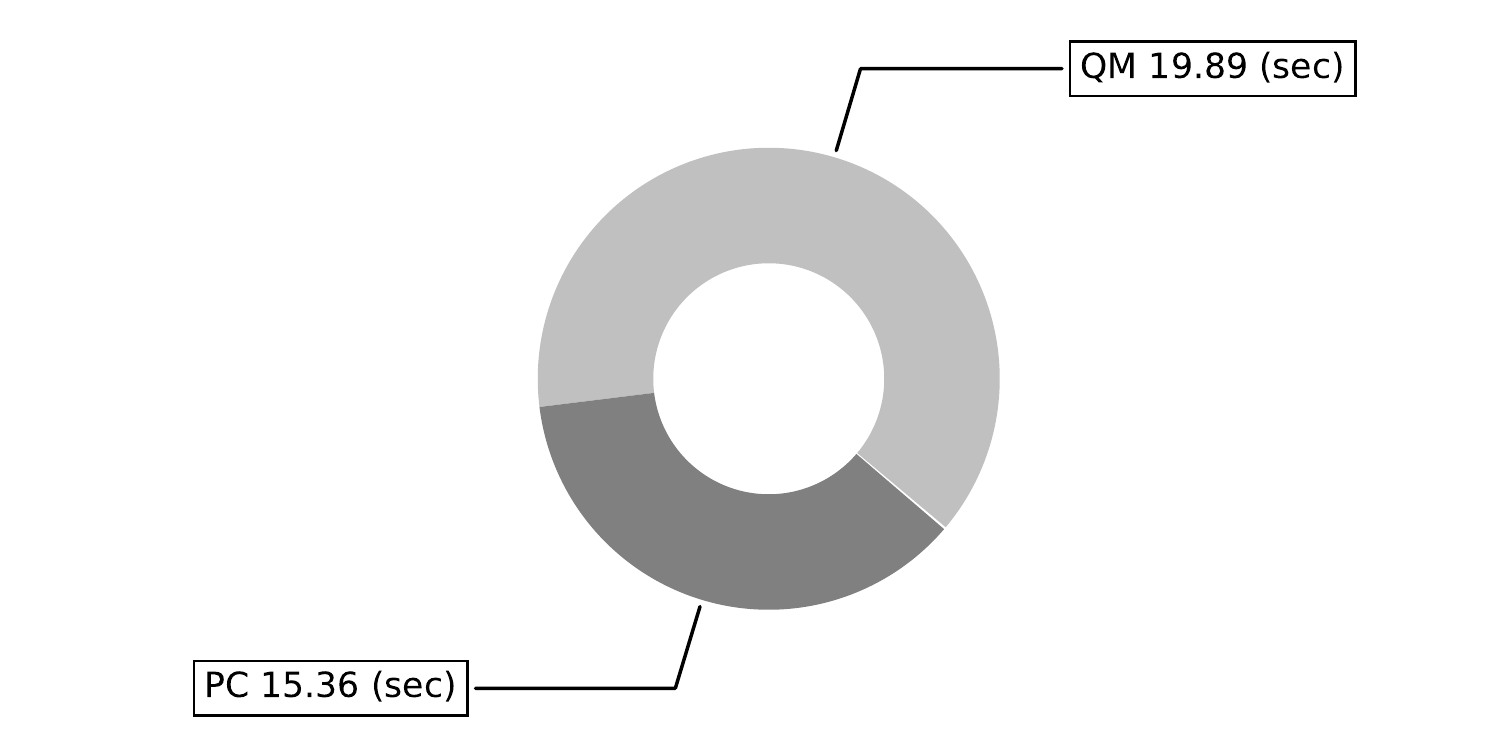}
	\caption{\emph{Piggybacked} with \emph{vt1-vt1}}
	\label{fig:piggy_vt1vt1}
\end{subfigure}
\begin{subfigure}{0.32\textwidth}
	\includegraphics[scale=0.45]{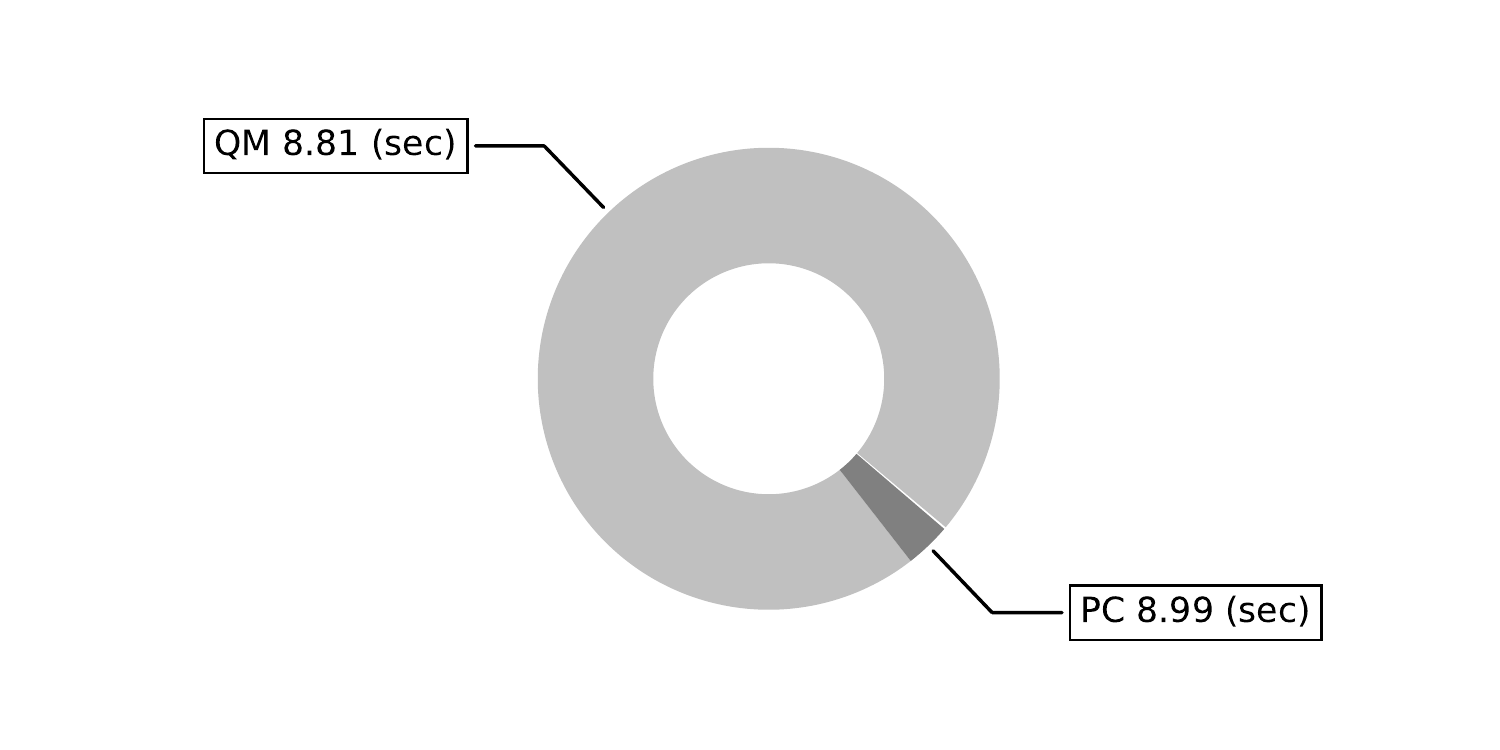}
	\caption{\emph{Piggybacked} with \emph{vt50p-vt50p}}
	\label{fig:piggy_vt50pvt50p}
\end{subfigure}
\begin{subfigure}{0.32\textwidth}
	\includegraphics[scale=0.45]{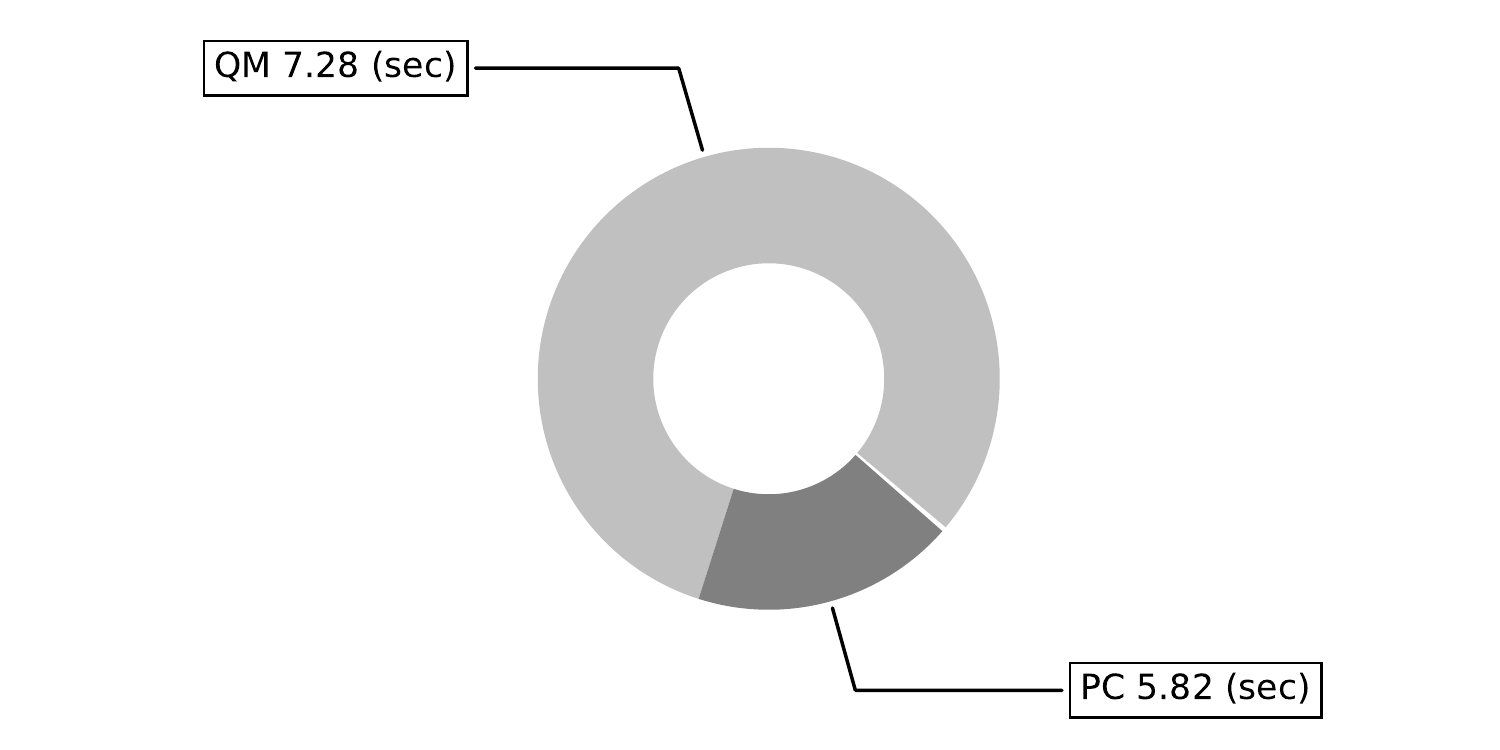}
	\caption{\emph{Piggybacked} with \emph{vt50p-vt1}}
	\label{fig:piggy_vt50pvt1}
\end{subfigure}
\begin{subfigure}{0.32\textwidth}
	\includegraphics[scale=0.45]{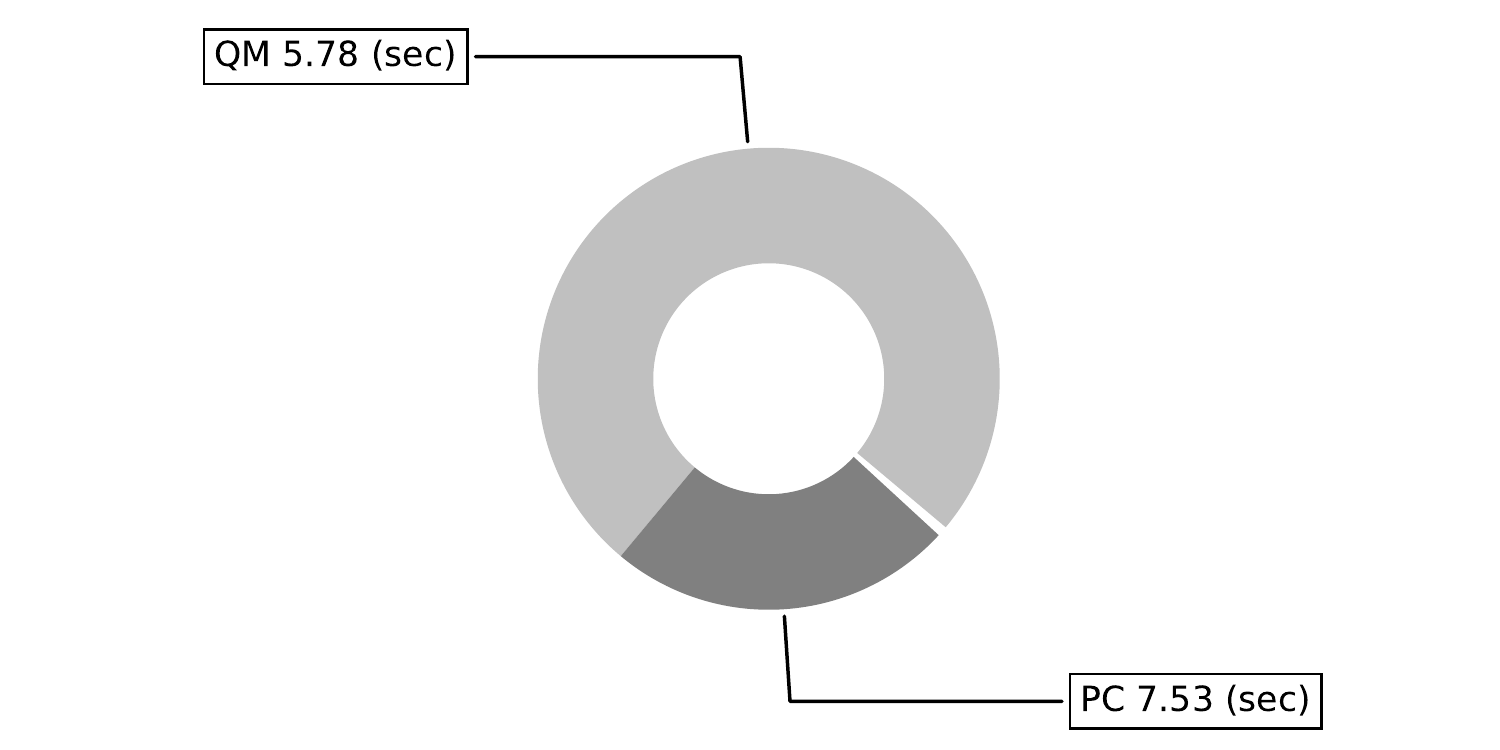}
	\caption{\emph{Original} with \emph{vt1-vt1}}
	\label{fig:org_vt1vt1}
\end{subfigure}
\begin{subfigure}{0.32\textwidth}
	\includegraphics[scale=0.45]{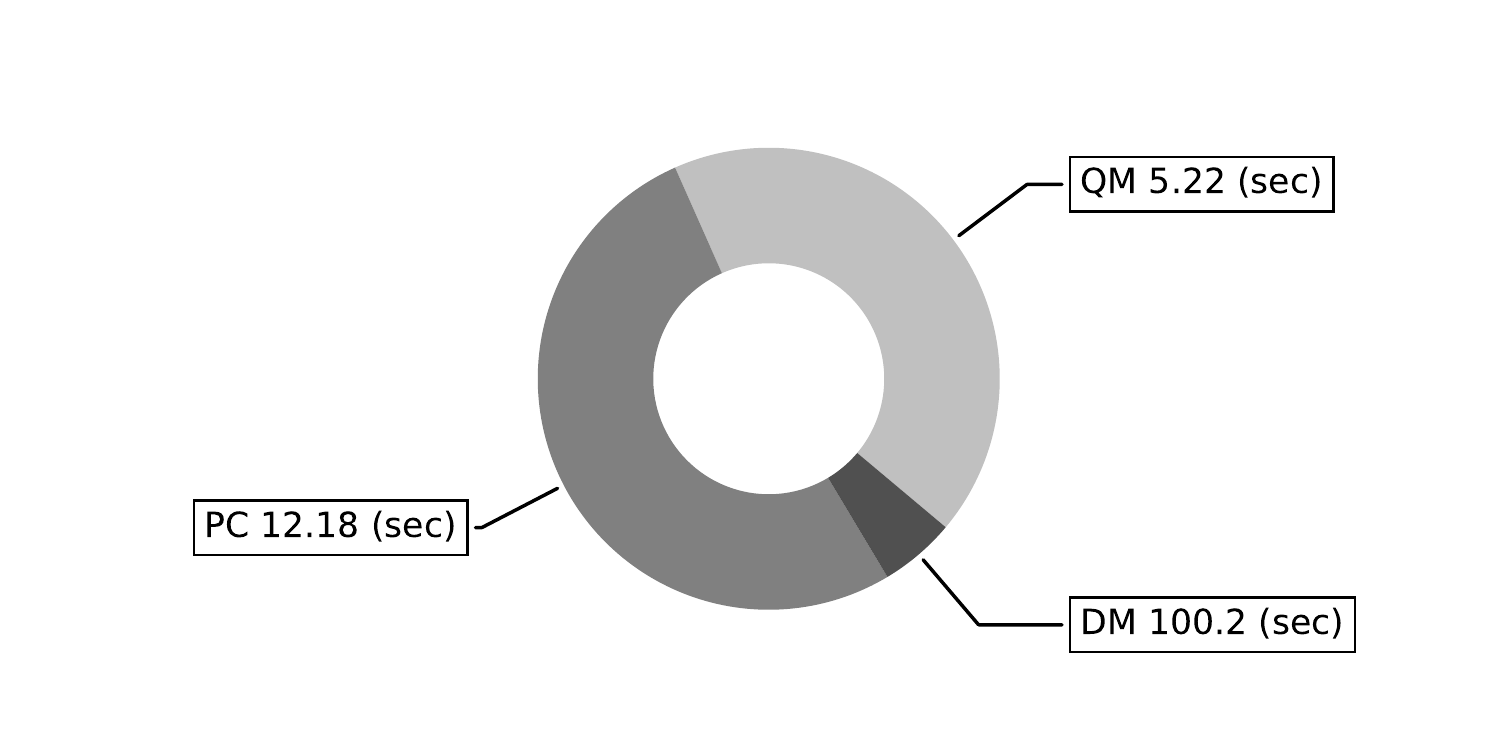}
	\caption{\emph{Original} with \emph{vt50p-vt50p}}
	\label{fig:org_vt50pvt50p}
\end{subfigure}
\begin{subfigure}{0.32\textwidth}
	\includegraphics[scale=0.45]{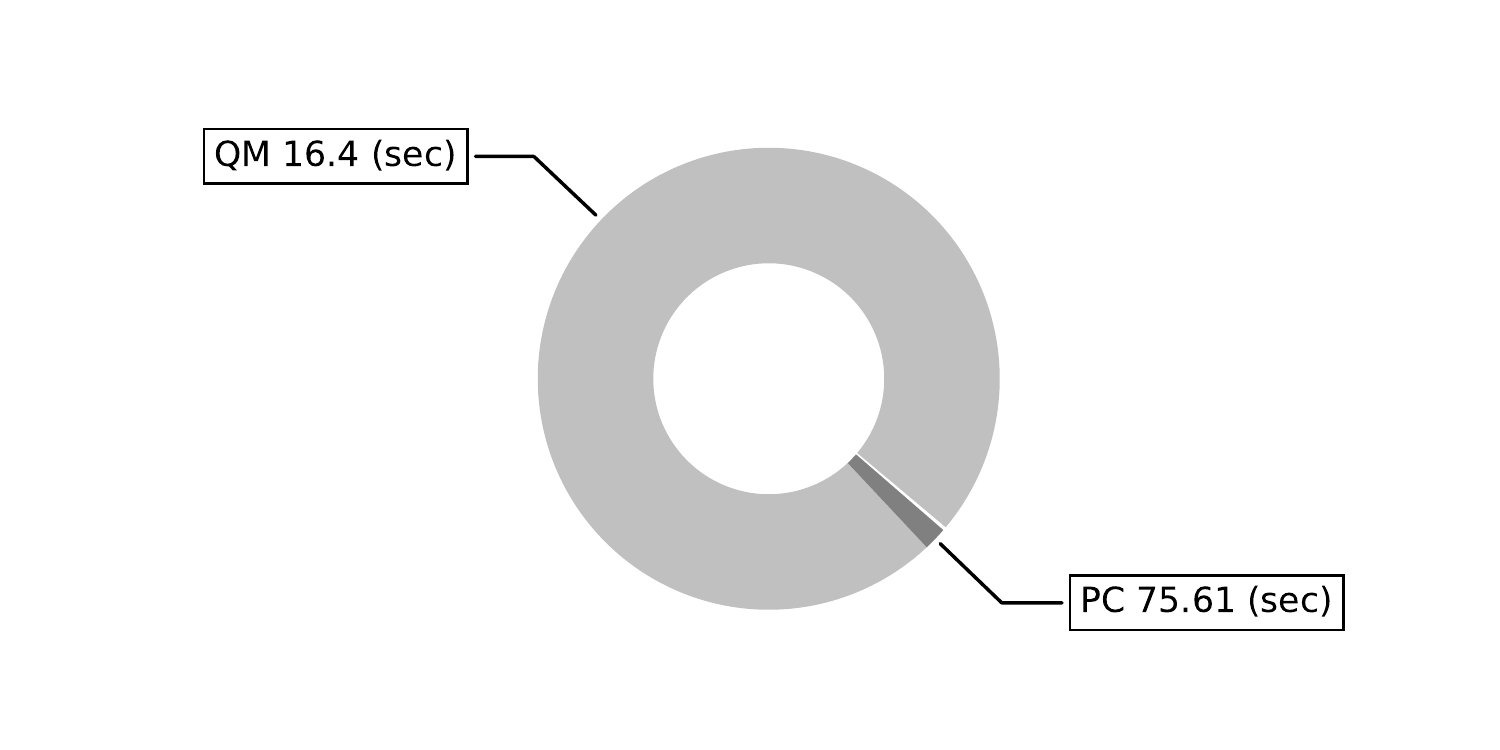}
	\caption{\emph{Original} with \emph{vt50p-vt1}}
	\label{fig:org_vt50pvt1}
\end{subfigure}
\begin{subfigure}{0.32\textwidth}
	\includegraphics[scale=0.45]{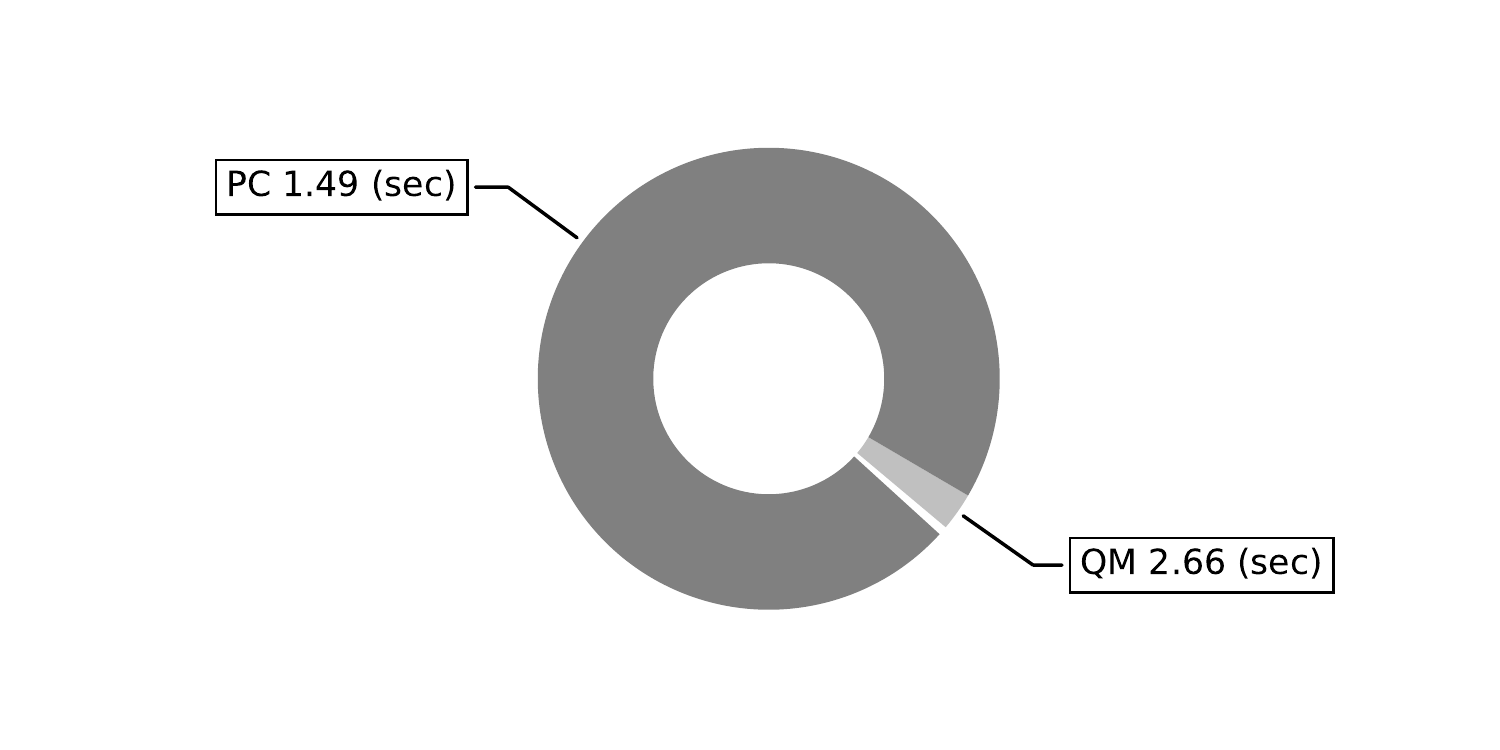}
	\caption{\emph{Malgenome} with \emph{vt1-vt1}}
	\label{fig:mal_vt1vt1}
\end{subfigure}
\begin{subfigure}{0.32\textwidth}
	\includegraphics[scale=0.45]{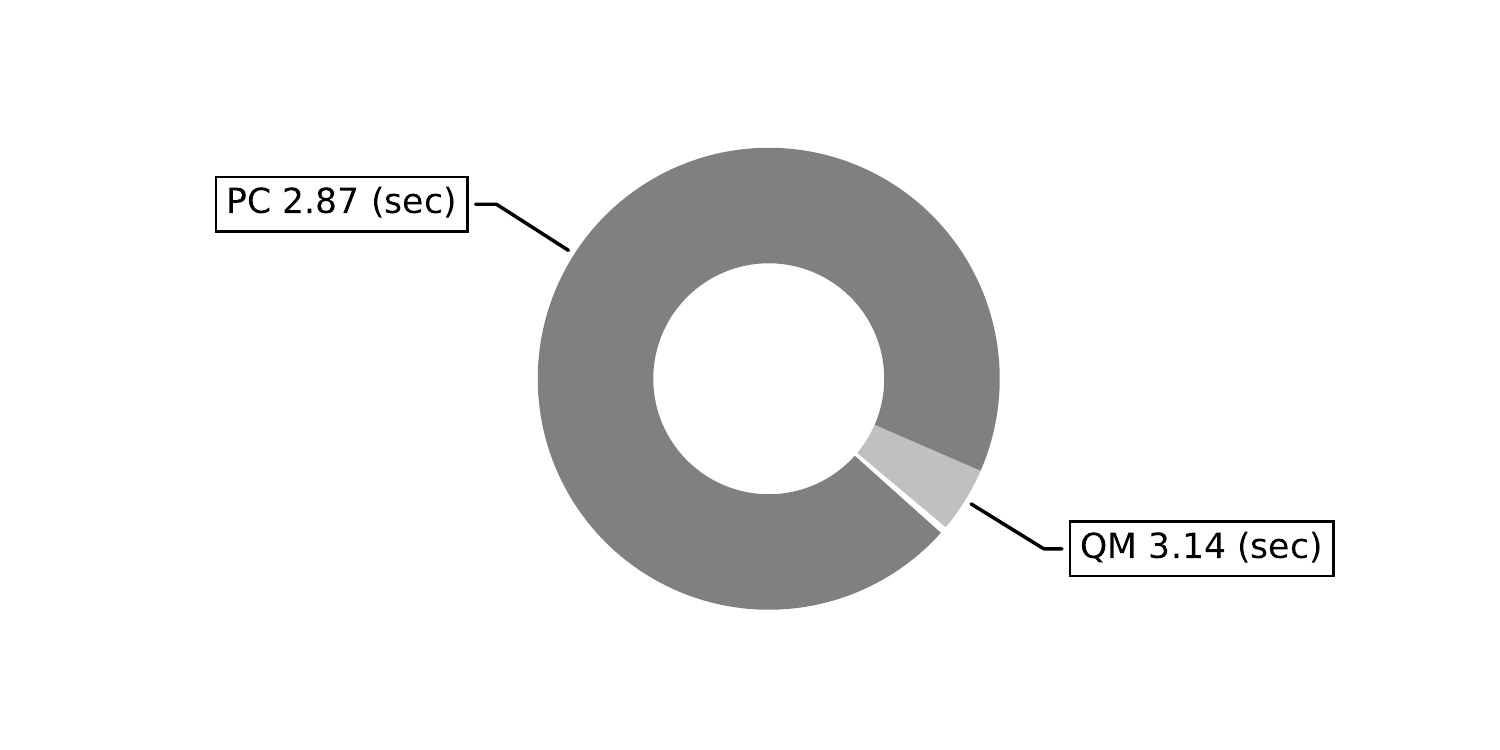}
	\caption{\emph{Malgenome} with \emph{vt50p-vt50p}}
	\label{fig:mal_vt50pvt50p}
\end{subfigure}
\begin{subfigure}{0.32\textwidth}
	\includegraphics[scale=0.45]{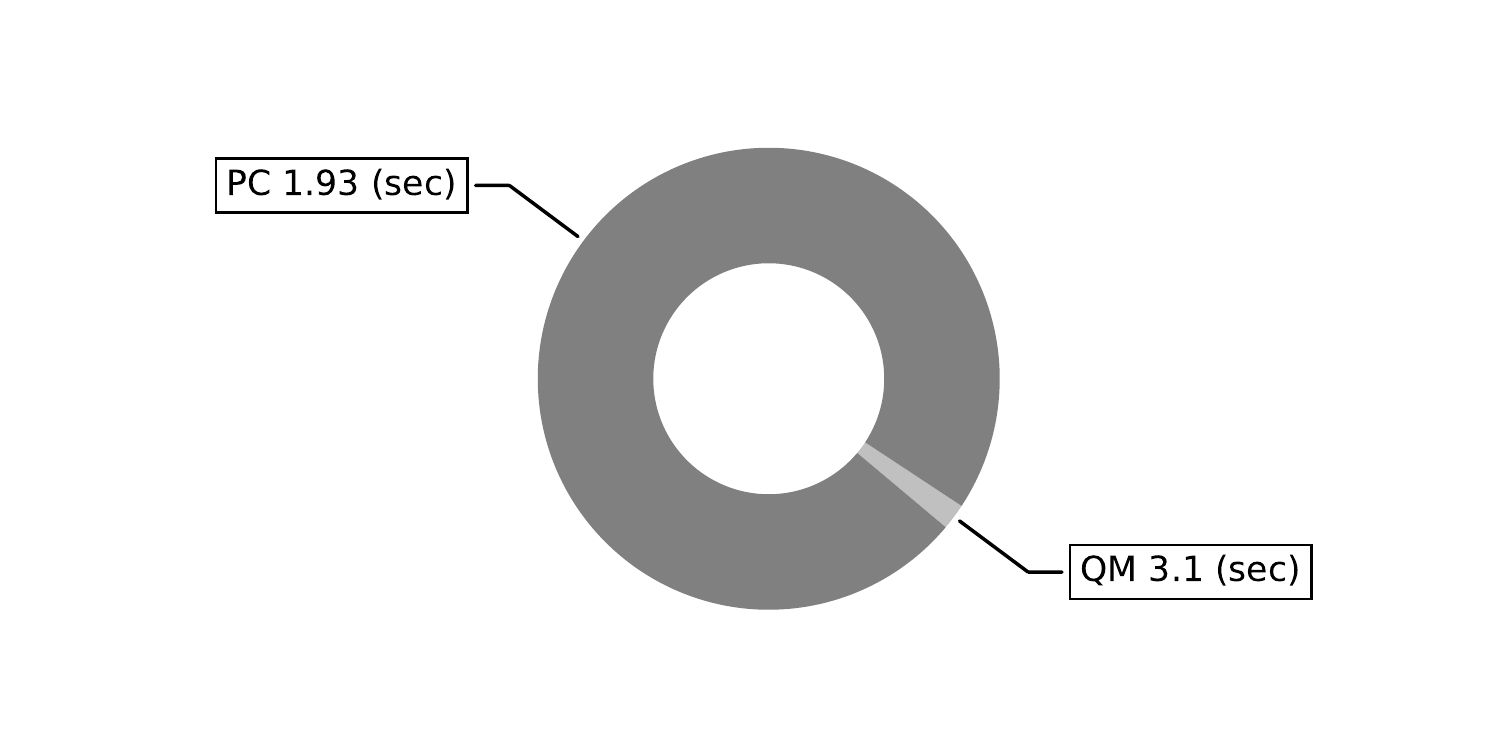}
	\caption{\emph{Malgenome} with \emph{vt50p-vt1}}
	\label{fig:mal_vt50pvt1}
\end{subfigure}
\caption{The contribution of each detection method to the amount of apps correctly classified by \dejavu's ensemble.}
\label{fig:pies}
\end{sidewaysfigure*}

In this section, we plot the contribution of each individual detection method to the detection accuracies of \dejavu's ensemble. 
Table \ref{tab:pies} summarizes such contributions in terms of the percentage of apps each detection method correctly classified as part of the ensemble for different dataset-labeling scheme combinations. 
For example, under the \emph{original} labeling scheme, 72\% of the apps correctly classified by \dejavu's ensemble were, in fact, classified by the \emph{quick matching} detection method with an average of 28.48 seconds taken to classify those apps.
We present this data in the form of pie charts in figure \ref{fig:pies}.

\end{document}